\documentclass[aps,
            pra,
            reprint,
            secnumarabic, 
            nofootinbib,longbibliography]{revtex4-2}
\usepackage{preamble}
\UseRawInputEncoding
\begin{document}

\title{Qudit vs. Qubit: Simulated performance of error correction codes in higher dimensions}
\author{James Keppens\,\orcidlink{0000-0001-5698-9549}$^{1,2}$}
\email{james.keppens@imec.be}
\author{Quinten Eggerickx\,\orcidlink{0000-0002-4709-3115}$^{1,2,3}$}
\author{Vukan Levajac\,\orcidlink{0000-0002-6985-822X}$^{1,4}$}
\author{George Simion\,\orcidlink{0000-0002-6880-6161}$^{1}$}
\author{Bart Sor\'ee\,\orcidlink{0000-0002-4157-1956}$^{1,2,5}$}

\affiliation{$^1$Imec, Leuven, Belgium}
\affiliation{$^2$Department of Electrical Engineering, KU Leuven, Leuven, Belgium}
\affiliation{$^3$Department of Physics, UGent, Ghent, Belgium}
\affiliation{$^4$Department of Physics, KU Leuven, Leuven, Belgium}
\affiliation{$^5$Department of Physics, Universiteit Antwerpen, Antwerpen, Belgium}

\date{\today}
 
\begin{abstract}
Qudits can be described by a state vector in a $q$-dimensional Hilbert space, enabling a more extensive encoding and manipulation of information compared to qubits. This implies that conducting fault-tolerant quantum computations using qudits rather than qubits might entail less overhead. In this work, we investigate the viability of qudits in error correction codes by creating and simulating the quantum circuitry for the smallest qudit error correction code with a multidimensional circuit-level noise model and specifically adapted decoders. After introducing a flag qudit to protect the code from hook errors, comparable error thresholds of the order of $10^{-4}$ are obtained for qudits of dimensions $2$, $3$ and $5$.
\end{abstract}
 
\maketitle 

\section{Introduction}
One of the main obstacles in advancing practical quantum computing lies in the significant susceptibility of quantum systems to noise \cite{Zurek_1991}. At present, the most promising approach to address this issue is through the utilization of error correction codes, an approach recently proven to work effectively in experiments conducted by Google Quantum AI and collaborators \cite{Acharya2024}. These codes combine $n$ physical qubits to construct $k$ logical units of which the logical states are separated by a distance $d$. Today's most practical code, the surface code \cite{Bravyi_Kitaev_Landau_1998, Fowler_Mariantoni_Martinis_Cleland_2012}, suffers from a sub-optimal scaling between $n$ and $d$. Additionally, it lacks an efficient method to perform certain logical gates, necessitating an excessive number of qubits to achieve satisfactory logical error levels for the execution of quantum algorithms. Efforts to address these problems involve research into linear high-distance Low-Density-Parity-Check (LDPC) codes \cite{Panteleev_Kalachev_2022,Bravyi_Cross_Gambetta_Maslov_Rall_Yoder_2024}, and Non-Abelian codes \cite{Field_Simula_2018,Schotte_Zhu_Burgelman_Verstraete_2022}. While these methods show promise, the majority of proposed codes either remain confined to acting solely as quantum memories or still entail excessive overhead requirements. Another approach to mitigate the overhead associated with quantum error correction codes involves the utilization of multilevel systems.\\
Qudits are $q$-level quantum mechanical systems, capable of storing and processing the same amount of information as $\log_2(q)$ qubits. Moreover, their extended complete gate set governed by $\bigotimes SU(q)$ allows for some more fundamental theoretical advantages \cite{Wang_Hu_Sanders_Kais_2020, Huber_de_Vicente_2013, Gokhale_Baker_Duckering_Brown_Brown_Chong_2019}. Leveraging qudits for fault-tolerant schemes can thus offer improved scaling options and enhanced methods to perform logical gates \cite{Campbell_Anwar_Browne_2012,Campbell_2014}, provided that the fabrication and control of qudits are not overly complex. These advantages sparked the interest in qudit basic control and led to several proposals and demonstrations in photonic systems \cite{Chi_etal._2022,Lanyon_Barbieri_Almeida_Jennewein_Ralph_Resch_Pryde_O’Brien_Gilchrist_White_2009}, superconducting circuits \cite{Bianchetti_Filipp_Baur_Fink_Lang_Steffen_Boissonneault_Blais_Wallraff_2010,Kononenko_Yurtalan_Ren_Shi_Ashhab_Lupascu_2021}, spin systems \cite{deFuentes_etal._2023,Godfrin_Ferhat_Ballou_Klyatskaya_Ruben_Wernsdorfer_Balestro_2017}, trapped ions \cite{Ringbauer_Meth_Postler_Stricker_Blatt_Schindler_Monz_2022} and Rydberg atoms \cite{Weggemans_Urech_Rausch_Spreeuw_Boucherie_Schreck_Schoutens_Minář_Speelman_2022,Ahn_Weinacht_Bucksbaum_2000}.\\
In this work, we investigate the use of qudits in quantum error correction codes under circuit-level noise. Our goal is to assess whether higher-dimensional qudits exhibit a practical error threshold that is within reach of near-term devices, and consequently evaluate their viability for fault-tolerant quantum computation. The essential theoretical background on higher-dimensional quantum computing and the stabilizer formalism is briefly summarized in Section \ref{sec: theory}. In Section \ref{sec: pqc}, we introduce the $5$-qudit code which is smallest stabilizer code capable of correcting any single qudit error \cite{Laflamme_Miquel_Paz_Zurek_1996} and derive a general quantum circuit for it, functional for qudits of all prime dimensions. In Section \ref{sec: decoders}, we present two widely recognized decoders for quantum error correction and modify them for applicability to higher-dimensional error correction codes. Furthermore, in Section \ref{sec: Simulations}, we describe the conducted simulations and multidimensional noise channels. After validating the adapted decoders under standard depolarizing noise, we introduce an additional flag qudit to enable the correction of hook errors in the presence of circuit-level noise. Finally, in Section~\ref{sec: Threshold}, we use concatenation to estimate the error thresholds of the $5$-qudit codes of dimensions $2$, $3$, and $5$.

\section{Multilevel fault tolerant quantum computing: Theory} \label{sec: theory}
\subsection{Higher dimensional quantum computing}\label{sec: HDQC}
For a $q$-dimensional qudit, basis states are indexed by an alphabet of size $q$, with the general state represented as $\ket{\psi} = \sum_{i=0}^{q-1} \alpha_i \ket{i}$, where $\alpha_i \in \mathbb{C}$ and $\sum_{i=0}^{q-1} |\alpha_i|^2 = 1$. A universal quantum gate set of dimension $q$ is defined as a collection of matrices $U_k \in U(q^n)$ whereby the collective multiplication of its elements enables the approximation of any arbitrary unitary transformation $U$ within the Hilbert space $\mathcal{H}^{\otimes n}_q$ \cite{Wang_Hu_Sanders_Kais_2020}. Numerous propositions of universal gate sets have been proposed \cite{Muthukrishnan_Stroud_2000,Brennen_O’Leary_Bullock_2005,Luo_Wang_2014}, yet the most pragmatic option to use within the stabilizer formalism is a set that contains the Clifford group and the Heisenberg-Weyl gates as a subset. The definitions of these gates in general dimension $q$ are given by the following expressions:
\begin{equation}
    X_q :=  \sum_{n=0}^{q-1}\ket{n+1}\bra{n}\quad ,
\end{equation}
\begin{equation}
    Z_q :=  \sum_{n=0}^{q-1}\omega^{n}\ket{n}\bra{n}  \quad ,
\end{equation}
\begin{equation}
\begin{aligned}
S_q &=
\begin{cases}
Z^{1 + \frac{q}{2}} P_q, & \text{if } q \text{ is even}, \\
Z^{\frac{1 + q}{2}} P_q, & \text{if } q \text{ is odd},
\end{cases} \\
P_q &:= \sum_{n=0}^{q-1} \omega^{\frac{n(n - q - 2)}{2}} \ket{n} \bra{n}
\quad, 
\end{aligned}
\end{equation}
\begin{equation}
    F_q := \sum_{n=0}^{q-1}\sum_{m=0}^{q-1}\frac{\omega^{nm}}{\sqrt{q}}\ket{m}\bra{n} \quad ,
\end{equation}
\begin{equation}
    SUM := \sum_{n=0}^{q-1}\sum_{m=0}^{q-1}\ket{n}\ket{m+n}\bra{n}\bra{m} \quad ,
\end{equation}
with all numbers taken modulo $q$ and $\omega = e^{\frac{2\pi i}{q}}$ the $q$'th root of unity. The operators $X_q$ and $Z_q$ are the generalized Pauli operators, also known as the Heisenberg--Weyl operators. Together with the generalized phase gate $S_q$, the Fourier transform gate $F_q$, and the modular addition gate $\mathrm{SUM}_q$, they generate the multi-qudit Clifford group up to a global phase factor. These gates satisfy the following conjugation relations \cite{Moussa_2016}:
\begin{equation}\label{eq: conj}
\begin{aligned}
Z_q X_q Z_q^\dagger &= \omega X_q, \\
F_q X_q F_q^\dagger &= Z_q, \\
F_q Z_q F_q^\dagger &= X_q^\dagger, \\
S_q X_q^\dagger S_q &= -\omega^{-1/2} X_q Z_q=Y_q, \\
S_q Z_q S_q^\dagger &= Z_q, \\
\mathrm{SUM}_q (X_q \otimes I) \mathrm{SUM}_q^\dagger &= X_q \otimes X_q, \\
\mathrm{SUM}_q (I \otimes X_q) \mathrm{SUM}_q^\dagger &= I \otimes X_q, \\
\mathrm{SUM}_q (Z_q \otimes I) \mathrm{SUM}_q^\dagger &= Z_q \otimes I, \\
\mathrm{SUM}_q (I \otimes Z_q) \mathrm{SUM}_q^\dagger &= Z_q^\dagger \otimes Z_q.
\end{aligned}
\end{equation}
This subset can be completed to a full universal gate set by adding a Non-clifford gate, such as a higher dimensional analog of the Toffoli gate \cite{Gottesman_1998}. An additional useful qudit gate is the multiplication gate: 
\begin{equation}
    M_{q} :=  \ket{n(q-1)}\bra{n} \quad ,
\end{equation}
in which the initial state is transformed via multiplication by $q-1$, with the result taken modulo $q$.

\subsection{Stabilizer formalism in higher dimensions}
A stabilizer code in $q$ dimensions is characterized by an Abelian subgroup $\mathcal{S}$ of $\mathcal{P}_q^{\otimes n}$ with $-I$ excluded \cite{Gottesman_1997,Gottesman_1998}. The codespace is spanned by the $+1$-eigenspace of $\mathcal{S}$ and detectable errors originating from a particular error basis lead to an exit from the codespace. For the codes regarded in this paper, the correctable error sets are composed by the single qudit Pauli group $\mathcal{P}_q$:
\begin{equation}\label{eq: error set}
    \mathcal{E}_q=\{\omega^aX_q^rZ_q^s\} \text{ with } 0\leq a,r,s < q 
\end{equation}
 and $\omega^a$ a phase factor accounting for the different commutations that satisfy the relation:
\begin{equation}\label{eq: relation errors}
    (X_q^r Z_q^s) (X_q^t Z_q^u) = \omega^{st - ru} (X_q^t Z_q^u) (X_q^r Z_q^s) \quad .
\end{equation}
These errors possess orthogonal eigenspaces with eigenvalues $\omega^a$, which are $q$'th roots of unity, and can therefore be interpreted as measurable and distinguishable physical observables. An $[[n,k,d]]_q$ stabilizer code of distance $d$ then has $n$ physical qudits representing $k$ logical qudits, generated by $n-k$ generators, a feature only true for prime dimension $q$ \cite{Gottesman_1998}.\\
Furthermore, any qudit stabilizer code has a valid multi-qudit transversal operation \cite{Gottesman_1998,Gottesman_qubit}. This operation, together with measuring entangled Pauli operators allows for the construction of a fault-tolerant logical $SUM$ gate and consequently the entire logical Clifford group. The fault tolerant logical gate set can then be completed using a higher dimensional analog of the Toffoli gate \cite{Gottesman_1998}.

\section{Perfect qudit codes}\label{sec: pqc}
The smallest error correction code that can correct any single qudit error contains 5 data qudits. This number is determined by the quantum Hamming bound (qHb) and quantum Singleton bound (qSb)\cite{Knill_Laflamme_1997,Gottesman_1997}. \\
The qHb is a condition generated by simply counting the amount of orthogonal subspaces required to accommodate all errors. In qudit codes with prime dimension, there are $q$ logical states that each need their own $q^2-1$ subspaces for all possible single error combinations of $X_q$ and $Z_q$, and one additional subspace for the unperturbed state. This creates a generalized quantum Hamming bound that prime dimension qudit codes need to satisfy:
\begin{equation}\label{eq: qHb}
    q\left(\left(q^2-1\right)n+1\right)\leq 
    q^n \quad .
\end{equation}
The qSb for an $[[n,k,d]]_q$ code is established by the fact that $n-(d-1)$ qudits must contain enough information to be able to reconstruct the $q^k$ logical states and the missing $d-1$ qudits:
\begin{equation}\label{eq: qSb}
    2(d-1)+k\leq n \quad .
\end{equation}
The smallest number that meets both conditions is $n=5$, which makes error correction codes with 5 data qudits the so called "perfect" codes. This section describes the construction of the complete quantum circuit for these codes.

\begin{figure*}[t]
    \centering
\begin{quantikz} \label{fig: circuit}
\lstick{$\ket{Q_0}$} & \gate{F_q^{\dagger}} & \qw & \qw & \qw & \qw & \qw & \qw & \qw & \qw & \ctrl{1} & \ctrl{2} & \ctrl{4} & \qw & \qw \\
\lstick{$\ket{Q_1}$} & \gate{F_q^{\dagger}} & \qw & \qw & \qw & \qw & \qw & \ctrl{1} & \ctrl{2} & \gate{M_q^{\dagger}} & \gate{+} & \qw & \qw &\gate{M_q^{\dagger}} & \qw \\
\lstick{$\ket{Q_2}$} & \gate{F_q^{\dagger}} & \qw & \qw & \ctrl{1} & \ctrl{2} & \gate{M_q^{\dagger}} & \gate{+}& \qw & \gate{F_q^{\dagger}} & \qw &\gate{+} & \qw & \gate{F_q^{\dagger}} & \qw \\
\lstick{$\ket{Q_3}$} & \gate{F_q^{\dagger}} & \ctrl{1} & \gate{M_q^{\dagger}}& \gate{+}  & \qw  & \gate{M_q^{\dagger}} & \qw & \gate{+} & \gate{F_q^{\dagger}} & \qw & \qw & \qw & \qw  & \qw\\
\lstick{$\ket{Q_4}$} & \qw & \gate{+} & \gate{F_q^{\dagger}}& \qw & \gate{+} & \gate{M_q^{\dagger}} & \qw & \qw & \qw & \qw & \qw & \gate{+} & \gate{M_q^{\dagger}} & \gate{F_q^{\dagger}}
\end{quantikz}
    \caption{Encoding circuit of the 5 qudit code, valid for all prime dimensions $q$.}
    \label{fig:encoding circ}
\end{figure*}

\subsection{The encoding circuit}
The stabilizer generators of these 5-qudit codes are cyclic permutations of $IZ_qX_qX_q^\dagger Z_q^\dagger$. The encoding circuitry can be created by transforming the parity check matrix of the code into one that represents the stabilizer of $\ket{00000}$ with the conjugation relations (\ref{eq: conj}) mentioned in  Section \ref{sec: HDQC} \cite{Grassl_Otteler_Beth_2002}. Starting from the general parity check matrix $(Z|X)$,
\[
\left(
\begin{array}{ccccc|ccccc}
  0 & 0 & q-1 & 0 & 1 & 1 & q-1 & 0 & 0 & 0 \\
  1 & 0 & 0 & q-1 & 0 & 0 & 1 & q-1 & 0 & 0 \\
  0 & 1 & 0 & 0 & q-1 & 0 & 0 & 1 & q-1 & 0 \\
  q-1 & 0 & 1 & 0 & 0 & 0 & 0 & 0 & 1 & q-1 \\
\end{array}
\right)
\]  and applying the following sequence of gates
\begin{equation*}
    G_1 = I \otimes M_{q} \otimes F_q \otimes I \otimes M_{q}F_q \quad ,
\end{equation*}
\begin{equation*}
    G_2 = SUM^{(0,1)\dagger}SUM^{(0,2)\dagger}SUM^{(0,4)\dagger} \quad ,
\end{equation*}
\begin{equation*}
    G_3 = I \otimes M_{q} \otimes F_q \otimes F_q \otimes I \quad ,
\end{equation*}
\begin{equation*}
    G_4 = SUM^{(1,2)\dagger}SUM^{(1,3)\dagger} \quad ,
\end{equation*}
\begin{equation*}
    G_5 = I \otimes I \otimes M_{q} \otimes F_q \otimes M_{q} \quad ,
\end{equation*}
\begin{equation*}
    G_6 = SUM^{(2,3)\dagger}SUM^{(2,4)\dagger} \quad ,
\end{equation*}
\begin{equation*}
    G_7 = I \otimes I \otimes I \otimes M_{q} \otimes F_q \quad ,
\end{equation*}
\begin{equation*}
    G_8 = SUM^{(3,4)\dagger} \quad ,
\end{equation*}
\begin{equation*}
    G_9 =  F_q \otimes F_q \otimes F_q \otimes F_q \otimes I \quad ,
\end{equation*}
results in the check matrix:
\[
\left(
\begin{array}{ccccc|ccccc}
  1 & 0 & 0 & 0 & 0 & 0 & 0 & 0 & 0 & 0 \\
  0 & 1 & 0 & 0 & 0 & 0 & 0 & 0 & 0 & 0 \\
  0 & 0 & 1 & 0 & 0 & 0 & 0 & 0 & 0 & 0 \\
  0 & 0 & 0 & 1 & 0 & 0 & 0 & 0 & 0 & 0 \\
\end{array}
\right)
.\] 
Inverting the sequence of gates and using their conjugate transpose versions, results in a general encoding circuit for perfect qudit codes. The quantum circuitry can be seen on Figure \ref{fig:encoding circ}.

\begin{figure*}[ht]
    \centering
    \resizebox{\textwidth}{!}{ 
    \begin{quantikz}\label{fig: circuit syndrome}
\lstick{$\ket{A_0}$}&\gate{F_q}& \ctrl{4} &\ctrl{5}  &\ctrl{6} &\ctrl{8} &  \qw & \qw & \qw & \qw & \qw & \qw &\qw &\qw & \qw &\qw & \qw & \qw &\gate{F_q^{\dagger}} & \gate{M}\\
\lstick{$\ket{A_1}$}&\gate{F_q}&\qw &\qw &\qw &\qw &\ctrl{4} &\ctrl{5}  &\ctrl{6} &\ctrl{3} & \qw &\qw & \qw & \qw & \qw &\qw & \qw & \qw&\gate{F_q^{\dagger}}& \gate{M}\\
\lstick{$\ket{A_2}$}&\gate{F_q}&\qw &\qw &\qw &\qw & \qw & \qw & \qw & \qw & \ctrl{4} &\ctrl{5}  &\ctrl{6} &\ctrl{3}&\qw &\qw & \qw & \qw &\gate{F_q^{\dagger}}& \gate{M}\\
\lstick{$\ket{A_3}$}&\gate{F_q}&\qw &\qw &\qw &\qw & \qw & \qw & \qw & \qw &\qw & \qw & \qw & \qw & \ctrl{4} &\ctrl{5}  &\ctrl{1} &\ctrl{3}&\gate{F_q^{\dagger}}& \gate{M}\\
\lstick{$\ket{Q_0}$\ldots}& \qw & \gate{+} &\qw &\qw &\qw &\gate{F_q^{\dagger}} & \qw & \qw & \gate{+} & \gate{F_q}& \qw & \qw & \qw & \gate{F_q^{\dagger}}  &\qw & \gate{+^{\dagger}} & \qw& \gate{F_q}&\qw\\
\lstick{$\ket{Q_1}$\ldots}&\qw &\qw &\gate{+^{\dagger}} &\qw &\qw & \gate{+} & \qw & \qw &\qw & \gate{F_q^{\dagger}}& \qw & \qw &\gate{+} & \gate{F_q} & \qw & \qw & \qw &\qw&\qw\\
\lstick{$\ket{Q_2}$\ldots}&\gate{F_q^{\dagger}}&\qw &\qw & \gate{+^{\dagger}} &\qw &\gate{F_q} & \gate{+^{\dagger}} & \qw &\qw &  \gate{+}& \qw &\qw & \qw & \gate{F_q^{\dagger}}  & \qw & \qw &\gate{+}& \gate{F_q}&\qw\\
\lstick{$\ket{Q_3}$\ldots}&\qw&\qw &\qw &\qw &\qw & \gate{F_q^{\dagger}} & \qw & \gate{+^{\dagger}} &\qw & \gate{F_q}&  \gate{+^{\dagger}}& \qw &\qw & \gate{+} & \qw & \qw &\qw&\qw&\qw\\
\lstick{$\ket{Q_4}$\ldots}&\gate{F_q^{\dagger}}&\qw &\qw &\qw &\gate{+}&\gate{F_q}&\qw &\qw &\qw&\gate{F_q^{\dagger}} & \qw &  \gate{+^{\dagger}} &\qw & \gate{F_q}& \gate{+^{\dagger}} & \qw &\qw&\qw&\qw\\
\end{quantikz}}
    \caption{Quantum circuit representing a single cycle of error syndromes of the 5 qudit code, valid for all prime dimensions $q$. This circuit is not optimized for minimal circuit depth. It is designed to present the structure in a clear manner. This circuit can be optimized by initializing and measuring the ancilla qudits immediately before and after their use, thereby eliminating idle periods. Additionally, a greater degree of parallelism can be applied to operations on the data qudits than depicted here, further reducing idle time. $Q$ represents a data qudit and $A$ an ancilla qudit. The SUM gates have a dot at the control qudit and a $+$ sign at the target qudit.}
    \label{fig: syndrome_circ}
\end{figure*}

\subsection{Syndrome measurements}
The next part of the quantum circuit should contain the parity checks of the stabilizers. 4 ancilla qudits are added to the circuitry to enable error syndromes that measure the eigenvalues of the stabilizer operators with projection operators. In this way, errors are discretized after each syndrome measurement. The quantum circuit for a single cycle of such an error syndrome can be seen on Figure \ref{fig: syndrome_circ}. At a later stage, this circuit is optimized to fit as many qudit gates as possible into each time step, minimizing idling time. It is important to note that for a code utilizing $N$ ancilla units, there are $q^N$ possible syndromes, which makes the use of a simple lookup table impractical and creates the necessity of using more efficient decoding procedures in the form of decoding algorithms.

\subsection{Fault tolerant computation}
Although the $5$-qudit code is optimal in its use of data qudits, as a non-CSS code it has fewer automorphisms (physical operations preserving stabilizers) than most quantum error correction codes. The single-qudit transversal operations for this code are $X_L= X_qX_qX_qX_qX_q$, $Z_L = Z_qZ_qZ_qZ_qZ_q$ and $T_L = T_qT_qT_qT_qT_q$, with $T_q=S_qF_q$  being the operation that cyclically shifts between the Heisenberg-Weyl gates. Additionally, we can construct the following three-qudit gate that acts on logical qudits:
\begin{equation}\label{eq: T3gate}
\begin{aligned}
X_q \otimes I_q \otimes I_q &\rightarrow X_q \otimes Y_q \otimes Z_q \\
I_q \otimes X_q \otimes I_q &\rightarrow Y_q \otimes X_q \otimes Z_q \\
I_q \otimes I_q \otimes X_q &\rightarrow X_q \otimes X_q \otimes X_q \\
Z_q \otimes I_q \otimes I_q &\rightarrow Z_q \otimes X_q \otimes Y_q \\
I_q \otimes Z_q \otimes I_q &\rightarrow X_q \otimes Z_q \otimes Y_q \\
I_q \otimes I_q \otimes Z_q &\rightarrow Z_q \otimes Z_q \otimes Z_q \quad .
\end{aligned}
\end{equation}
This gate can be implemented transversally using the qudit quantum circuit shown in Figure \ref{fig: T3gate}. When combined with transversal Pauli measurements and the preparation of stabilizer states, it enables the realization of any logical gate within the Clifford group \cite{Cross_DiVincenzo_Terhal_2009}. Note that this method is a natural extension of the one mentioned for qubits in \cite{Gottesman_1997,Cross_DiVincenzo_Terhal_2009}.
\begin{figure*}[ht]
    \centering
    \resizebox{\textwidth}{!}{ 
    \begin{quantikz}\label{fig: T3gate}
    \lstick{$\ket{L_2}$} & \gate{Y_q} & \gate{(S_qF_qS_qF_q)^\dagger} & \gate{S_q^\dagger} & \gate{+} & \gate{S_q} & \gate{(S_qF_qS_qF_q)^\dagger} & \gate{S_q^\dagger} & \gate{+} & \qw & \ctrl{2} & \rstick{$\ket{L_1}$} \\
    \lstick{$\ket{L_3}$} & \qw & \qw & \qw & \qw & \qw & \qw & \qw & \ctrl{-1} & \gate{+} & \qw & \rstick{$\ket{L_2}$} \\
    \lstick{$\ket{L_1}$} & \gate{Y_q} & \qw & \qw & \ctrl{-2} & \gate{(S_qF_qS_qF_q)} & \gate{S_q^\dagger} & \qw & \qw & \ctrl{-1} & \gate{+} & \rstick{$\ket{L_3}$} \\
    \end{quantikz}}
    \caption{ Quantum circuit representing the encoded implementation of the transversal 3-qudit gate (\ref{eq: T3gate}). Each gate in this scheme is applied transversally on the physical qudits of the corresponding logical qudit.}
    \label{fig: T3gate}
\end{figure*}

\section{Decoders and adaptations to higher dimensions}\label{sec: decoders}
In a realistic application of an error correction code, measurement errors can pose considerable problems as they can generate illusory faults and obscure genuine operational errors. Therefore, it is crucial to execute several syndrome measurement cycles and track actual detection events, which are characterized by a change in the measured eigenvalues. More specifically, if the measurement outcome of an ancilla during cycle $c$ is $m_c$ and the outcome in the previous cycle is $m_{c-1}$, then a detection event in cycle $c$ occurs whenever $m_{c-1} \oplus_q m_{c} \neq 0$ ($m_{-1}$ is considered to be zero). To account for errors during the measurement extraction, we add the requirement of $m_c\oplus_q m_{c+1} = 0$ (except when $c$ is the last cycle). The decoder then uses this information and determines a single logical correction tensor to apply at the end of the computation, eliminating the necessity for potentially noisy corrections after each cycle.\\
There are multiple types of decoding algorithms, each with their own decoding fidelities and speeds. Two of them are described and adapted to qudits in this section.\\

\subsection{Minimum-weight perfect matching} \label{sec: mwpm}
\begin{figure}[h!htb]
    \centering \includegraphics[width=0.49\textwidth]{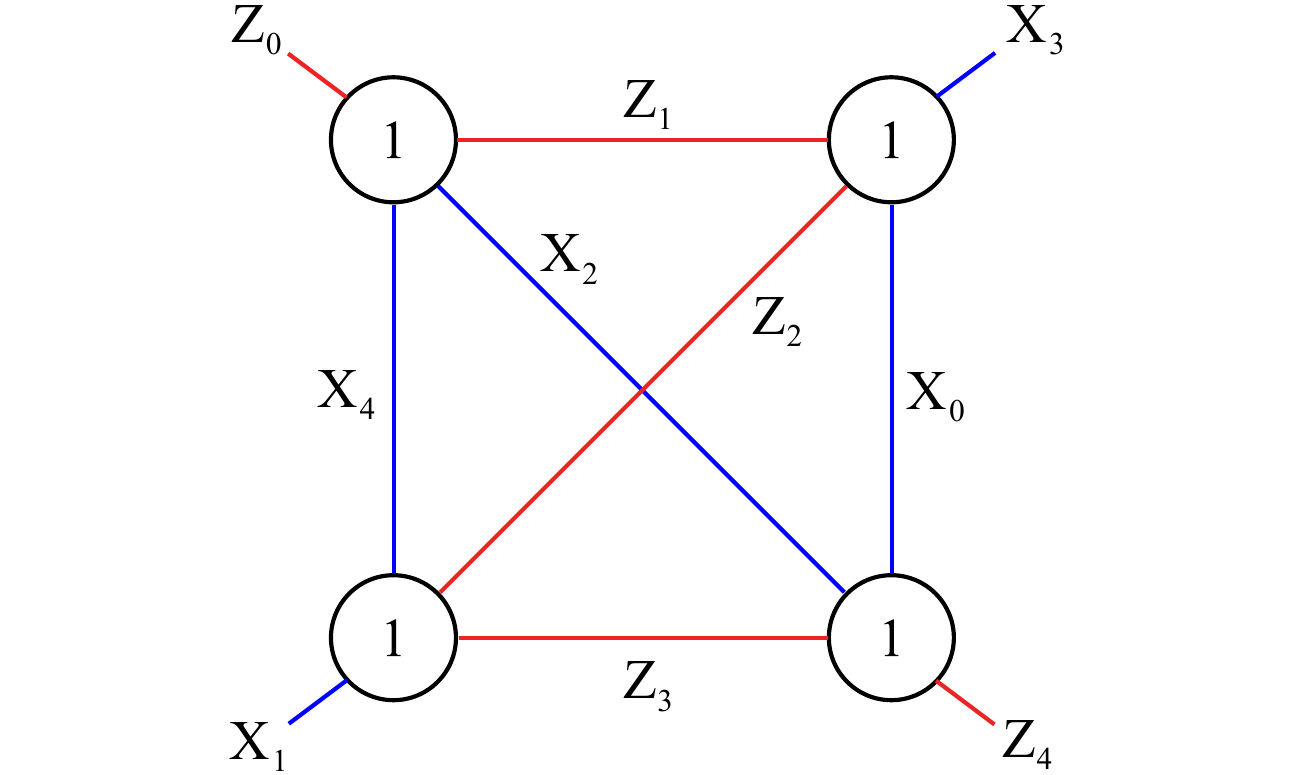}
    \caption{Matching graph of a single cycle 5 qubit code. All $X$ and $Z$ gates are $q=2$-dimensional and act on the $i$'th qudit with $i$ being the sub-scripted number. The $4$ nodes represent the ancilla qudits that are used to measure the eigenvalue of a stabilizer, which is able to detect the errors on the surrounding edges. Detection events identify specific nodes, which the decoder connects to determine the corresponding error.}
    \label{fig: MG2}
\end{figure}

\begin{figure}[!htb]
    \centering
    \includegraphics[width=0.49\textwidth]{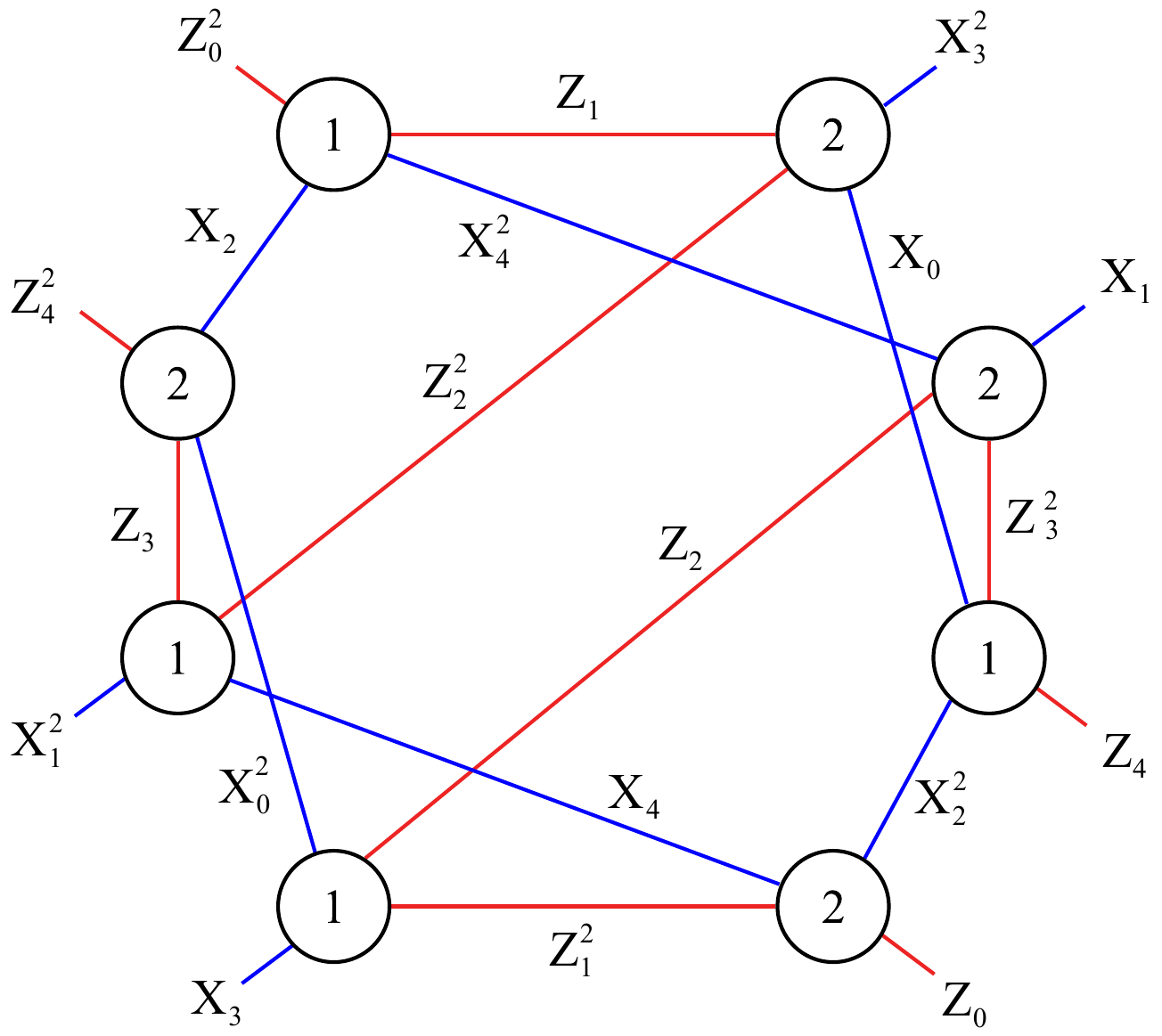}
    \caption{Matching graph of a single cycle 5 qutrit code. All $X$ and $Z$ gates are $q=3$-dimensional and act on the $i$'th qudit with $i$ being the sub-scripted number. $8$ nodes are now introduced, each representing one of the two measurable eigenvalues of the stabilizers. The edges surrounding each node correspond to the error that gives rise to the eigenvalue in question. Detection events identify specific nodes, which the decoder connects to determine the corresponding error.}
    \label{fig: MG3}
\end{figure}

\begin{figure*}[!htb]
    \centering
    \includegraphics[width=\textwidth]{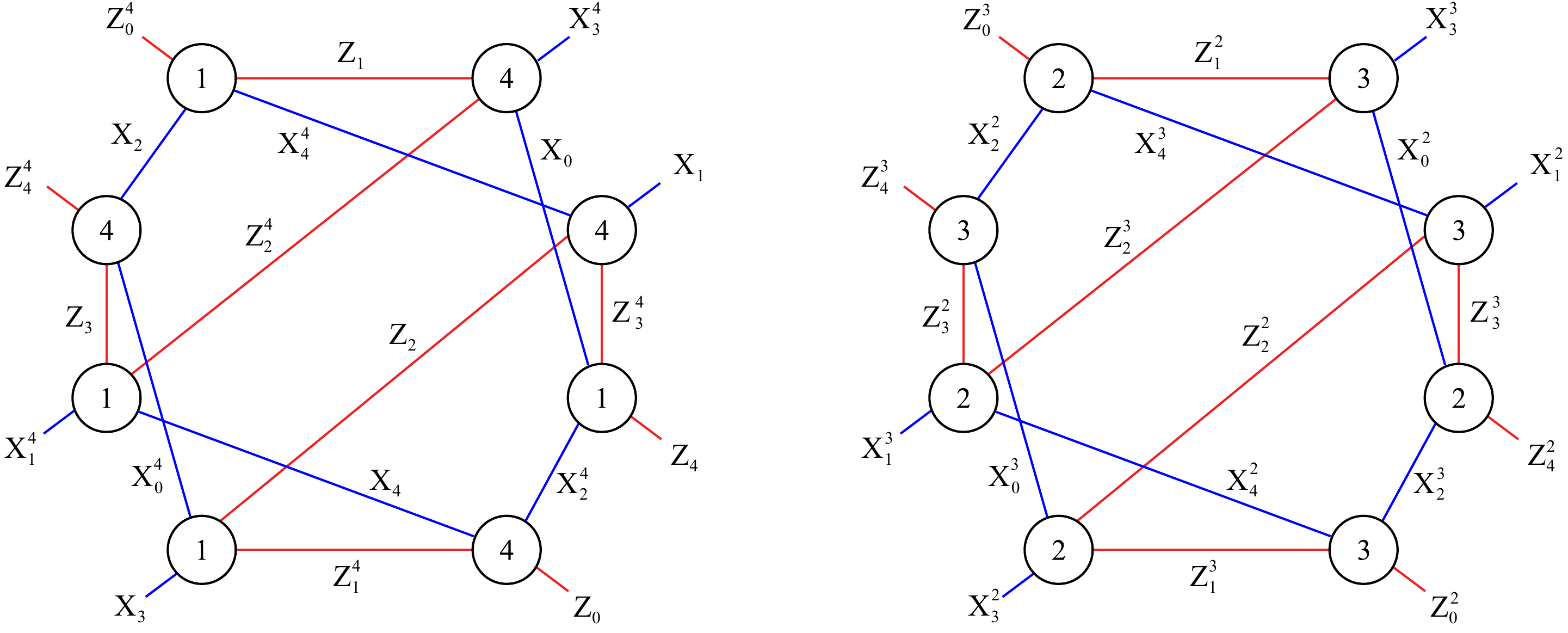}
    \caption{Matching graph of a single cycle 5 ququint ($q=5$) code. All $X$ and $Z$ gates are $q=5$-dimensional and act on the $i$'th qudit with $i$ being the sub-scripted number. This graph contains $4(q-1)=16$ nodes, each corresponding to one of the four distinct eigenvalues measurable by the four stabilizers. The edges surrounding each node correspond to the error that gives rise to the eigenvalue in question. Furthermore, the graph is disconnected and consists of 2 components with complementary errors and eigenvalues.}
    \label{fig: MG5}
\end{figure*}

The most commonly used and one of the fastest decoders for qubit error correction codes is the minimum-weight perfect matching (MWPM) algorithm \cite{Higgott_2022}. In this algorithm, a detector graph is constructed from the parity check matrix, the nodes represent the ancilla qubits, and each incident edge represents a bit or phase error that the corresponding ancilla can detect \cite{Antipov_Kiktenko_Fedorov_2023}. A fast way to find the minimum weight paths between detection events directly in the detector graph, is by using the sparse blossom algorithm, which has an expected running time that is linear in the amount of nodes in the graph \cite{Higgott_2022}\footnote{V2 \url{https://pymatching.readthedocs.io/en/stable/}}.\\
An adaptation to a version that can decode qudit codes can be achieved by creating the detector graph with $q-1$ nodes for every ancilla, representing all $q-1$ possible measurable eigenvalues and resulting in a total of $4(q-1)$ nodes for the $5$-qudit code. This method establishes a well-defined correspondence between a higher-dimensional parity check matrix and the associated detector graph, making it generally applicable to all qudit stabilizer error correction codes. A detailed explanation of this method is provided in Appendix~\ref{app: A}. An example of matching graphs for dimensions $2$, $3$ and $5$ can be seen on Figures \ref{fig: MG2}, \ref{fig: MG3} and \ref{fig: MG5}, respectively.

However, the disadvantage of this decoder is that it loses efficiency when hyperedge errors such as $Y_q$ are present in the noise model. These errors connect more than 2 nodes in the detector graph and should instead be represented on a hypergraph. Whenever a hyperedge can be decomposed into multiple equally weighted paths, the decoder's likelihood of identifying the right correction decreases. Each hyperedge error involves either three or four nodes in the matching graph. With a uniform weighting of edges, imposed for example by the depolarizing noise channel, the optimal solution consistently uses the minimum number of edges required to connect these nodes, which is two. In the qubit matching graph, depicted in Figure \ref{fig: MG2}, there exist three distinct configurations for connecting the nodes associated with a hyperedge error using two edges. An example of this is the $Y_0 = X_0Z_0$ error in Figure \ref{fig: MG2}. The three nodes connected by the edges of this error can alternatively be linked by two additional sets of two edges, namely $\{Z_1,Z_4\}$ or $\{X_2,X_3\}$. Conversely, in the qutrit ($q=3)$ matching graph (Figure \ref{fig: MG3}) there are only two possible configurations for connecting the nodes associated with a hyperedge error using two edges. This reduction in degeneracy results in a higher error correction fidelity for the MWPM decoder, as is demonstrated later by the results presented in Section \ref{sec: sdep}. At even higher dimensions ($q\geq5$), a clear pattern emerges. The matching graphs become disconnected, with each component resembling the structure observed in qutrit matching graphs, where two complementary eigenvalues are represented per component. Hyperedge errors that span edges in multiple components are efficiently corrected. However, those confined to a single component still suffer from the degeneracy problem. \\

\subsection{Belief matching}
An effective approach to address hyperedge errors during decoding is through the use of the belief propagation (BP) algorithm \cite{Criger_Ashraf_2018, roffe_decoding_2020}. This algorithm uses the parity check matrix $H$ and an initial error probability distribution $P(E)$ to create a Tanner graph and iteratively updates $P(E)$ by passing messages along the edges. The decoder also takes a syndrome $\mathbf{s}$ as input and stops whenever the syndrome equation $H\cdot E=\mathbf{s}$ is satisfied. This method ensures that hyperedge errors are consistently identified with a specified probability, rather than with a reduced probability as in the MWPM decoder. However, the algorithm does not account for measurement errors and lacks guaranteed convergence in the presence of cycles or multiple equally weighted solutions. To address these limitations, it is often combined with secondary codes that provide correction when belief propagation fails to converge. In this work we chose to use the Belief Matching (BM) algorithm that combines belief propagation together with the MWPM algorithm \cite{Higgott_Bohdanowicz_Kubica_Flammia_Campbell_2023}. \\
Adapting the decoder for higher-dimensional qudits is again a matter of adapting the input graph for the decoding algorithm to run on. We can use the same technique of adding detector nodes for the different possible eigenvalues of the stabilizers. In contrast to the MWPM decoder, belief propagation requires a prior error probability distribution, which can become challenging to generate for higher-dimensional systems, as the number of possible error configurations scales with $q^2$. 

\section{Simulations}\label{sec: Simulations}
\begin{figure*}[!htb]
    \centering    \includegraphics[width=0.95\textwidth]{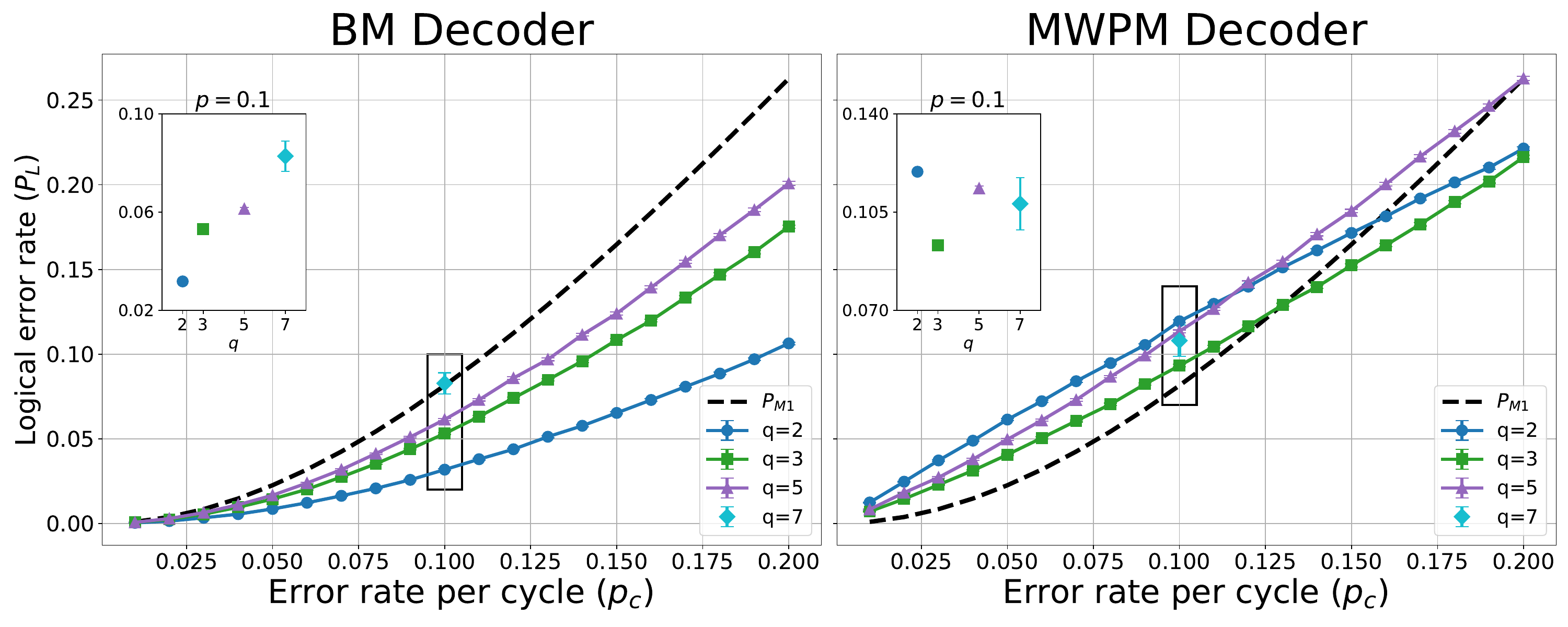}
    \caption{Logical error rates obtained in simulations with standard depolarization noise of the 3-cycle 5-qudit code using qudits of dimensions 2, 3, 5, and 7 decoded with the MWPM and BM decoders. The black dashed line represents the probability of experiencing at most a single qudit error per cycle, denoted by $P_{\text{M1}}$. The error bars correspond to the $99\%$ confidence intervals.}
    \label{fig: sdep}
\end{figure*}
To analyze how the qudit codes perform, their circuits were simulated with $\mathcal{O}(d)$ error syndrome cycles and two different noise models in the quantum circuit simulator Cirq \cite{Developers_2024}. Although Cirq is designed for qubits, it is extendable to simulate quantum circuits involving qudits by creating custom higher-dimensional gates. We therefore added several Python classes with unitary matrices representing the gates mentioned in section \ref{sec: HDQC}. The Python classes and a tutorial version of the simulations are available on GitHub \cite{Keppens2025}. \\
During the simulation, the extracted syndromes are input into the aforementioned decoders to gain a correction tensor $C$, which is applied to the state vector to create a residual error $\Gamma = CE$, with $E$ the actual error. The decoding is considered successful if the final state vector, $\Gamma\ket{\psi}_L$, matches the initial state vector up to the application of a stabilizer.

\subsection{Noise channels}
In the quantum circuit, noise is modeled using the operator-sum approach by introducing time slices with Kraus operators that evolve the density matrix via $\rho\longrightarrow\sum_iK_i\rho K_i^\dagger$. In this representation, the single qudit depolarization noise channel with Pauli operators is described by the mapping \cite{Wilde_2013}:
\begin{equation}\label{eq: Noisechannel}
    \rho \longrightarrow \left(1 - p\frac{\left(q^2 - 1\right)}{q^2}\right)\rho + \frac{p}{q^2}\sum_{i=1}^{q^2-1} P_i\rho P_i^\dagger \quad ,
\end{equation}
with $P_i$ one of the $q^{2}-1$ possible Pauli errors and $p$ the probability of having a depolarizing error. Note that the overall robustness of the qudit depends on the dimension as the chance of no error is $1 - p\left(q^2 - 1\right)/q^2$. Similarly, the two-qudit depolarizing channel is represented by the map:
\begin{multline}\label{Eq: noise2qudit}
    \rho \longrightarrow \left(1 - p_2\frac{\left(q^4 - 1\right)}{q^4}\right)\rho +\\
    \frac{p_2}{q^4} \left( \sum_{i=1}^{q^2-1}\sum_{j=1}^{q^2-1}  P_jP_i\rho P_i^\dagger P_j^\dagger + \sum_{i=1}^{q^2-1} P_i\rho P_i^\dagger + \sum_{j=1}^{q^2-1} P_j\rho P_j^\dagger \right) \quad ,
\end{multline}
with $P_i$ and $P_j$ one of the $q^{2}-1$ possible Pauli errors acting on two different qudits and $p_2$ the chance on a two-qudit depolarization error. Lastly, the measurement error channel is given by:
\begin{equation}
    \rho \longrightarrow \left(1 - p_m\right)\rho + p_m X_q\rho X_q^\dagger \quad .
\end{equation}
These noise channels are used to create two different noise models. \\
\subsection{Standard depolarization noise}\label{sec: sdep}
In the first model, single-qudit depolarization noise is applied to all data qudits prior to each cycle's syndrome extraction, while assuming perfect gates, perfect ancillas and ideal measurements with no errors. The input parameter $p_c$ corresponds to the error rate of a single physical qudit during one entire cycle. This simplified model is commonly used to benchmark error-correcting codes and decoders. Similarly, it is used here to demonstrate that the adapted decoders are effective for the higher-dimensional 5-qudit code.\\
The results of the simulations with standard depolarization noise can be seen in Figure \ref{fig: sdep}. The data points for dimensions $2$, $3$, and $5$ are based on averages computed from 400{,}000 simulated samples, ensuring a reliable convergence of the results. The error bars, representing $90\%$ confidence intervals, are smaller than the marker sizes for these dimensions. The black dashed line represents the probability of experiencing at most a single qudit error per cycle, denoted by $P_{M1} = 1 - \left[(1-p_c)^5+5p_c(1-p_c)^4)\right]$. Given that the $5$-qudit code has a code distance of three, it is theoretically capable of correcting any single error per cycle. Consequently, a fully functional decoder is expected to achieve a logical error rate $P_L$ that remains below $P_{M1}$. This is the case for the adapted BM decoder, making the dependence on the qudit dimension the sole factor influencing the logical error rate. This dependency is twofold. Firstly, the noise channel (\ref{eq: Noisechannel}) is inherently dependent on the system's dimensionality and secondly, as the dimension of the qudit increases, the likelihood that errors cancel each other out diminishes.\\
In contrast, using the less effective but faster MWPM decoder results in higher logical error rates compared to $P_{M1}$ as it does not always detect hyperedge errors. However, it is noteworthy that the decoder initially shows lower logical error rates for higher-dimensional codes than for the qubit version. This feature arises from the fact that hyperedge errors in higher dimensions can be decomposed into fewer, equally weighted paths, as was described in section \ref{sec: mwpm}. We additionally observe that for ququints ($q=5)$, a larger fraction of $Y$-type errors remains uncorrected compared to the qutrit case and at higher error probabilities the twofold influence of dimensionality in the noise channel becomes the dominant factor.\\
An additional data point for qudits of dimension $q=7$ at a physical error rate of $p=0.1$ is shown on the insets of Figure \ref{fig: sdep}. This data point is based on only 5000 simulated samples and is solely used to provide some confirmation of expected trends. More specifically, the BM-decoded data point shows a slightly higher logical error rate, which aligns with the expectation that error likelihood increases with dimensionality. For the MWPM-decoded data point, the logical error rate falls between those of qubits and qutrits. This behavior can be attributed to the structure of the matching graph, which is similar to that of the ququint case, consisting of a disconnected graph with three components. \\

\subsection{Circuit-level noise} \label{subsec: Cl noise}
\begin{figure*}[!htb]
    \centering    \includegraphics[width=0.99\textwidth]{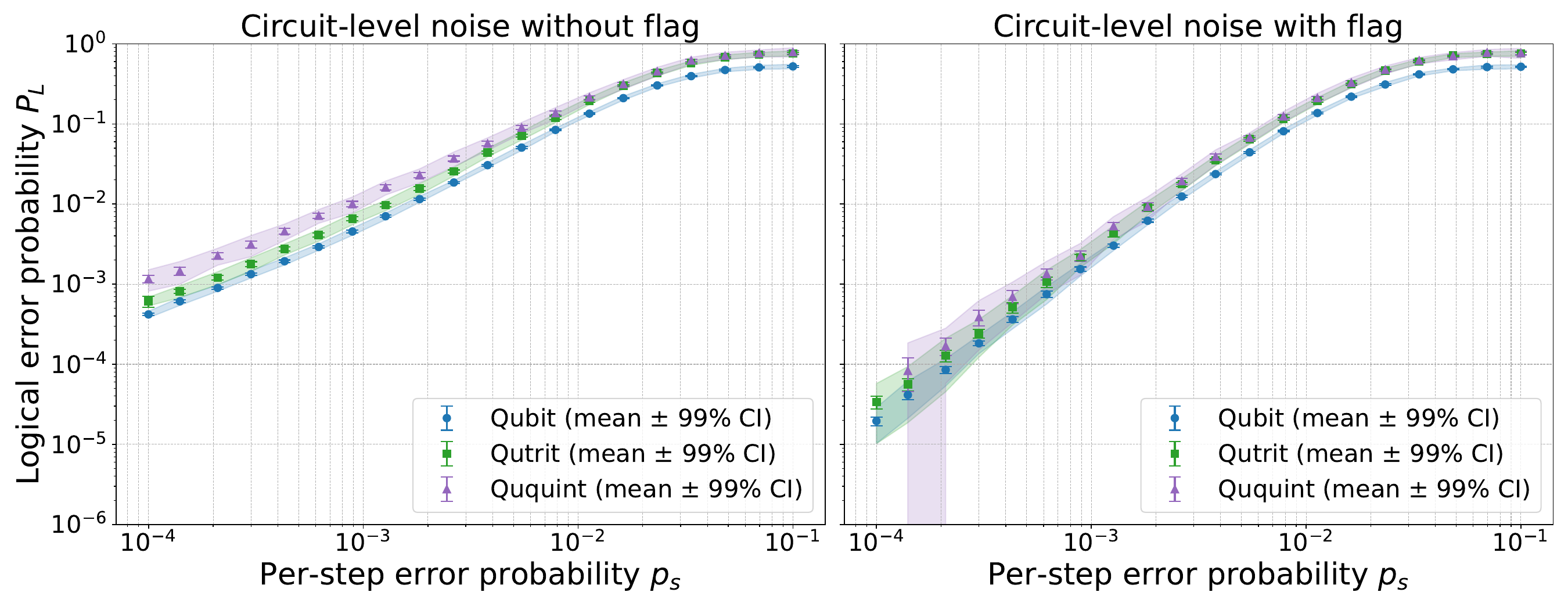}
    \caption{Simulated logical error probabilities for the 5-qudit code under circuit-level noise, using qudits of dimensions 2, 3, and 5. The left plot shows results without flag qudits, and the right plot shows results with flag qudits, both decoded using the BM decoder. Error bars indicate $90\%$ confidence intervals, and shaded regions represent one standard deviation.}
    \label{fig: real}
\end{figure*}

The second noise model is designed to better reflect the behavior of physical quantum circuits by incorporating errors at each timestep of the circuit with a per-step probability of $p_s$, following the convention of \cite{Fowler_Mariantoni_Martinis_Cleland_2012}. During idle periods, data qudits experience single-qudit depolarizing noise. For every  single-qudit gate operation, the respective qudit experiences single-qudit depolarizing noise. Conversely, when a qudit participates in a two-qudit gate operation, it is subject to two-qudit depolarizing noise. In both cases $p_s$ is plugged in the respective noise channel map, (\ref{eq: Noisechannel}) or (\ref{Eq: noise2qudit}), to account for the dimensionality and a symmetric distribution of Pauli errors. In addition, ancilla qudits are modeled to undergo bit-flip errors prior to their readout, except during the last syndrome extraction cycle, which is considered to be error-free. An important consequence of having errors on ancillas is that some of these errors are able to spread back to data qudits, creating the so called correlated hook errors. To remove these errors and keep the error correction code fault tolerant, we employ an extra flag qudit \cite{Chao_Reichardt_2020}. By coupling it to ancilla qudits for every stabilizer, as illustrated in the circuit shown in Figure \ref{fig: flag circuit}, the system can flag the occurrence of hook errors. Whenever that occurs, the additional information from the flag qudit is used to correct the hook error with a brute-force look-up table. The flag qudit is subject to two-qudit gate noise and bit-flip noise prior to its readout. Furthermore, the assumption is made that it can be measured and reinitialized fast enough to avoid introducing additional idle moments on data qudits \cite{Chao_Reichardt_2020}. In the optimized syndrome extraction circuit, each cycle consists of 10 time steps, during which each data qudit undergoes at most four idle periods.\\
\begin{figure}[!ht]
    \centering
    \resizebox{0.475\textwidth}{!}{ 
    \begin{quantikz}
    \lstick{$\ket{A_0}$}&\gate{F_q}& \ctrl{2} & \ctrl{1}&\ctrl{3}  &\ctrl{4}& \ctrl{1} &\ctrl{5}&\gate{F_q^{\dagger}} & \gate{M}\\
    \lstick{$\ket{F}$} & \qw & \qw  & \gate{+} & \qw  & \qw &\gate{+^{\dagger}}&\qw&\qw & \gate{M}\\
    \lstick{$\ket{Q_0}$\ldots}& \qw & \gate{+} & \qw &\qw &\qw &\qw & \qw &\qw&\qw\\
    \lstick{$\ket{Q_1}$\ldots}&\qw &\qw & \qw &\gate{+^{\dagger}} &\qw & \qw &\qw & \qw&\qw\\
    \lstick{$\ket{Q_2}$\ldots}&\gate{F_q^{\dagger}}&\qw & \qw &\qw & \gate{+^{\dagger}} & \qw &\qw &\gate{F_q} & \qw\\
    \lstick{$\ket{Q_4}$\ldots}&\gate{F_q^{\dagger}}&\qw &\qw &\qw & \qw & \qw &\gate{+}&\gate{F_q}&\qw \\
    \end{quantikz}}
    \caption{Example of a syndrome extraction quantum circuit with an additional flag qudit $\ket{F}$. The flag qudit is measured, reinitialized, and subsequently reused for the extraction of the next stabilizer in an identical manner.}
    \label{fig: flag circuit}
\end{figure}
Figure \ref{fig: real} shows simulated logical error rates for the 3-cycle 5-qudit error correction code under circuit-level noise. All data presented from this point onward are obtained using the BM decoder, which was selected based on its superior reliability as demonstrated in the preceding section. The left plot shows results without a flag qudit, while the right plot shows data from simulations that includes one to help correct hook errors. The data points represent the average logical error probability computed over 50 independent simulations, each consisting of $A/p_s$ samples. The parameter $A$ is assigned a value of $20$ for qubits, $5$ for qutrits, and $2$ for all data points corresponding to ququints with a flag qudit, except for the left-most point. For ququints without a flag qudit, as well as for the left-most point in the presence of a flag qudit, $A$ is set to $1$. These values are selected to balance computation time with statistical reliability, ensuring a sufficient number of logical errors for meaningful analysis. Error bars represent $99\%$ confidence intervals, while the shaded regions correspond to one standard deviation. A comparison of the two plots reveals that employing flag qudits to suppress hook errors not only decreases the overall logical error rate $P_L$, but also narrows the performance gap between qudits and qubits, suggesting that higher-dimensional qudits are more susceptible to hook errors. The fact that the logical error rates for higher-dimensional qudits are slightly worse than those for qubits is expected due to the nature of the noise models.

\subsection{Threshold estimation}\label{sec: Threshold}
The optimal five-qudit code is a unique quantum error-correcting code defined over the Galois Field $GF(4)$. Unlike topological codes such as the surface code, it does not exhibit a natural pathway for scalability. In particular, optimal codes with distances five and seven correspond to different unique code constructions \cite{Calderbank_Rains_Shor_Sloane_1997}. As a result, increasing the code distance for the $5$-qudit perfect code is only achievable through concatenation. \\ 
When the five-qudit code is concatenated with itself $l$ times, it yields a code with parameters $[[5^l, 1, 3^l]]$. To estimate a fault-tolerance threshold, simulations must be performed for at least three different distances, corresponding in this case to three levels of concatenation $l=1,2,3$. However, the computational cost of such large simulations with qudits is far beyond what is currently computationally feasible. To address this limitation, we employ a level-by-level approach to estimate the threshold \cite{Yoshida_Tamiya_Yamasaki_2025}. In this approach, the logical error rates of concatenation level $l$, simulated with circuit-level noise, are used as a new per-step error rate $p_s$ in concatenation level $l + 1$. Errors of all logical operations are thus estimated to have the same logical error rate. Instead of re-running simulations using the logical error rates from the previous concatenation level, we fit a power law to the simulated logical error rates in section \ref{subsec: Cl noise} and use it to project the logical error rates of subsequent concatenation levels. The fitting curve is given by:
\begin{equation} \label{eq: fit}
     P_L(p_s)=ap_s^b \quad,
\end{equation}
which reflects the probability of uncorrectable errors under the circuit-level noise model. The motivation behind this curve is that a distance-three quantum error-correcting code, such as the perfect code or the Steane code \cite{Yoshida_Tamiya_Yamasaki_2025}, is designed to correct a single independent physical error. Therefore, the leading contribution to the logical error rate arises from instances in which two independent errors occur, which, for sufficiently small physical error rates $p_s$, scale as $\sim p_s^2$. To account for both decoding errors and the fact that, in the absence of a flag qudit, hook errors remain uncorrectable, the power-law exponent $b$ is treated as a fit parameter rather than being fixed to the value 2. The values of the fits for all dimensions can be found in table \ref{tab: fits}.

\begin{table}[h!]
\centering
\caption{Error thresholds and fitting parameters $a$ and $b$ of the power-law model (Eq.~\ref{eq: fit}) for the logical error rate $P_L(p_s)$ for qubits, qutrits, and ququints with and without a flag qudit.}\label{tab: fits}
\begin{tabular}{@{}llccc@{}}
\toprule
Dimension & Flag & $a$ & $b$ & Threshold \\
\midrule
2 (qubit)     & no     & $36.7$  & $1.264$ &  $1.21 \times 10^{-6}$\\
2 (qubit)     & yes    & $766$   & $1.873$ & $4.95 \times 10^{-4}$\\
3 (qutrit)    & no     & $58.7$  & $1.288$ & $7.22 \times 10^{-7}$\\
3 (qutrit)    & yes    & $1116$  & $1.870$ & $3.24 \times 10^{-4}$\\
5 (ququint)   & no     & $35.3$  & $1.149$ & $4.36 \times 10^{-11}$\\
5 (ququint)   & yes    & $792$  & $1.798$ & $2.32 \times 10^{-4}$ \\
\bottomrule
\end{tabular}
\end{table}

\begin{figure*}[!htb]
    \centering    \includegraphics[width=0.99\textwidth]{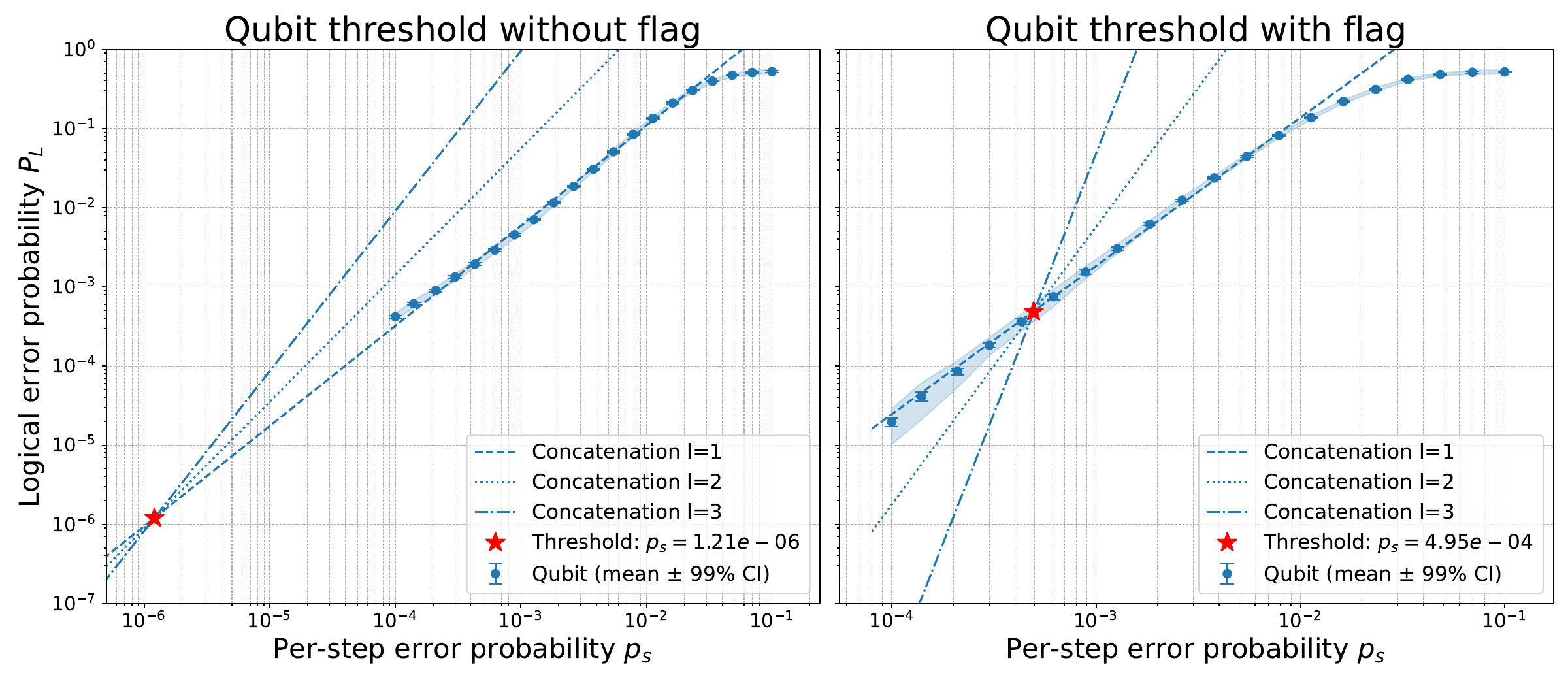}
    \caption{Simulated logical error probabilities for the $5$-qubit code decoded by the BM decoder with fitting curves for three concatenation levels. The left plot shows results without flag qudits, and the right plot shows results with the use of flag qudits. The error bars are $99\%$ confidence intervals, the shaded region represents a standard deviation and the star is the intersection point of the three fitting curves.}
    \label{fig: qubit threshold}
\end{figure*}

\begin{figure*}[!htb]
    \centering    \includegraphics[width=0.99\textwidth]{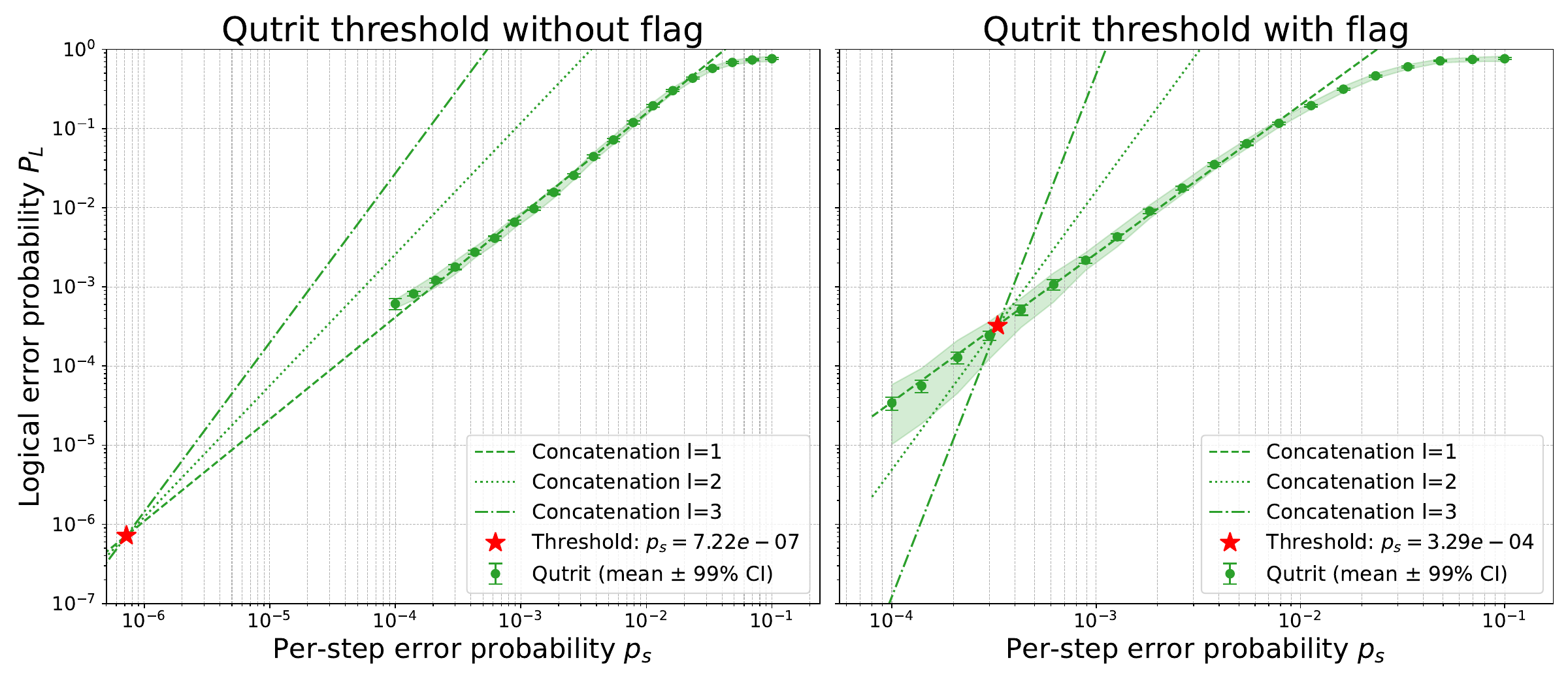}
    \caption{Simulated logical error probabilities for the $5$-qutrit code decoded by the BM decoder with fitting curves for three concatenation levels. The left plot shows results without flag qudits, and the right plot shows results with the use of flag qudits. The error bars are $99\%$ confidence intervals, the shaded region represents a standard deviation and the star is the intersection point of the three fitting curves.}
    \label{fig: qutrit threshold}
\end{figure*}

\begin{figure*}[!htb]
    \centering    \includegraphics[width=0.99\textwidth]{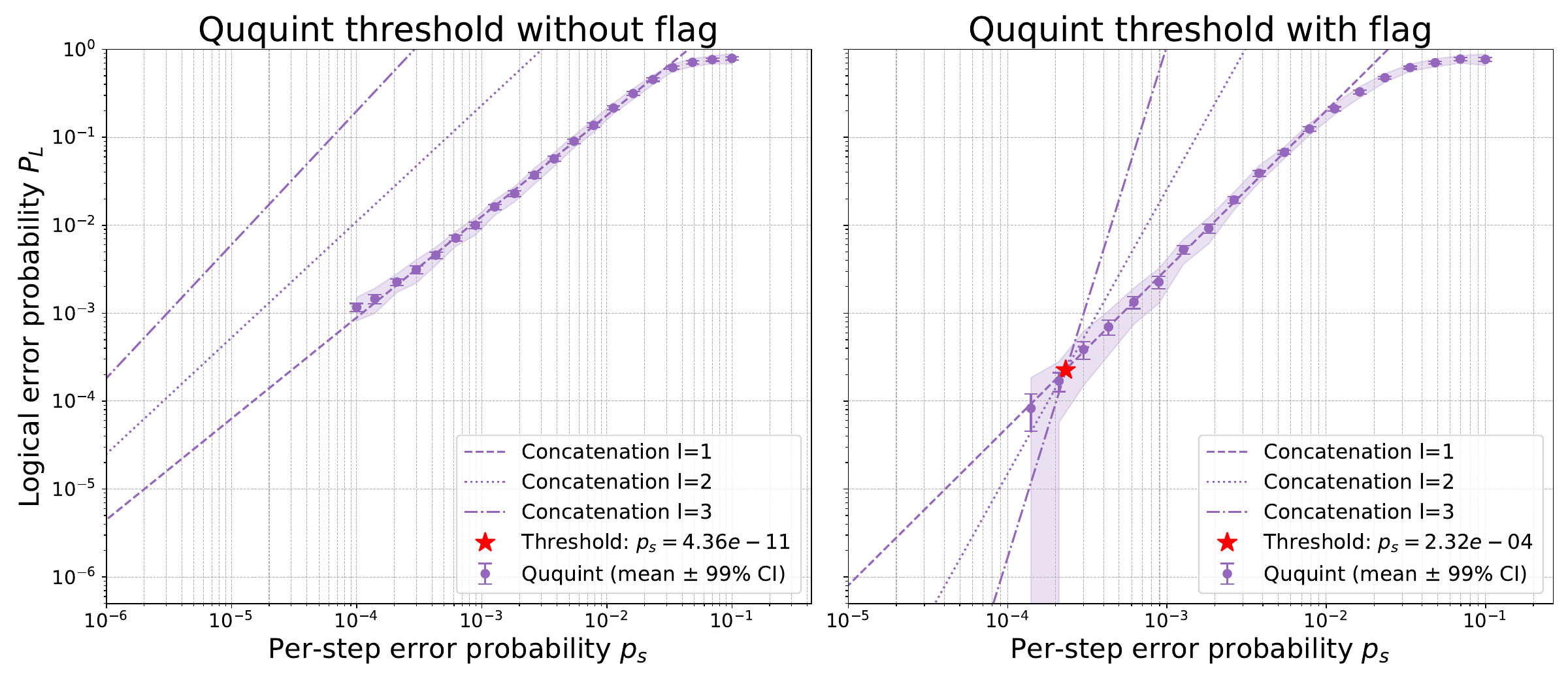}
    \caption{Simulated logical error probabilities for the $5$-ququint code decoded by the BM decoder with fitting curves for three concatenation levels. The left plot shows results without flag qudits, and the right plot shows results with the use of flag qudits. The error bars are $99\%$ confidence intervals, the shaded region represents a standard deviation and the star is the intersection point of the three fitting curves.}
    \label{fig: ququint threshold}
\end{figure*}

The curves for the subsequent concatenation levels on Figures \ref{fig: qubit threshold}, \ref{fig: qutrit threshold} and \ref{fig: ququint threshold} are then obtained by plugging in the obtained logical probability as a per-step probability for the next concatenated level: $P_L^{(l+1)} = P_L(p_s)\circ P_L^l$. By extrapolating these curves and determining the intersection point, we can get a good estimation of the threshold of these codes. The threshold values determined from simulations conducted without employing a flag qudit are as follows: $T^{(q=2)}_{\text{no flag}} = 1.21 \times 10^{-6} \pm 7.2 \times 10^{-8}$ for qubits, $T^{(q=3)}_{\text{no flag}} = 7.22 \times 10^{-7} \pm 7.6 \times 10^{-8}$ for qutrits, and $T^{(q=5)}_{\text{no flag}} = 4.36 \times 10^{-11} \pm 2.9 \times 10^{-11}$ for ququints. These thresholds are significantly improved when a flag qudit is utilized, yielding $T^{(q=2)}_{\text{flag}} = 4.95 \times 10^{-4} \pm 5.0 \times 10^{-6}$, $T^{(q=3)}_{\text{flag}} = 3.24 \times 10^{-4} \pm 6.5 \times 10^{-6}$ and $T^{(q=5)}_{\text{flag}} = 2.32 \times 10^{-4} \pm 1.0 \times 10^{-5}$. Uncertainties on the estimated error thresholds were obtained using a Monte Carlo propagation approach, sampling fit parameters from their covariance matrix and computing the resulting distribution of intersection points between the recursively applied power-law curves. Notably, qutrits and ququints now exhibit error thresholds comparable to those of qubits, which may (re)incentivize the use of qudits in quantum computations.

\section{Conclusion and outlook}
In this work, we have created a general encoding circuit for the $5$-qudit error correction code, proposed a general method to perform fault tolerant logical operations, performed simulations of syndrome measurements with two different noise models and provided evaluations of the error threshold under circuit-level noise. The extracted syndromes have been error corrected with adapted versions of the MWPM and BM decoder that work for any prime qudit dimension. Our analysis reveals that under the standard depolarization noise model, the adapted BM decoder consistently corrects all types of single qudit errors. The adapted MWPM decoder demonstrates lower performance compared to the BM decoder, primarily due to its reduced effectiveness in correcting hyperedge errors. Interestingly, this disadvantage becomes less pronounced at higher dimensions, as the larger matching graphs exhibit fewer degenerate solutions. Under circuit-level noise, ancilla qudits are susceptible to errors that can propagate as correlated hook errors, potentially compromising fault tolerance. To address this, we introduced an additional flag qudit to facilitate the identification and correction of such errors. The logical error probabilities decoded using the BM decoder revealed that hook errors were the primary factor limiting performance. Moreover, by fitting a power-law model to the data and applying a level-by-level concatenation method to estimate the thresholds, we found that the thresholds for qutrits and ququints were comparable to those of qubits. These results suggest that, despite being subject to an error model that scales with dimension, qudits remain a viable and promising platform for fault-tolerant quantum computation.\\
Future research directions include extending stabilizer simulation frameworks, such as STIM \cite{gidney2021stim}, to support qudit-based systems, which could enable the simulation and exploration of more complex and large-scale qudit error correction codes. Additionally, we propose adapting alternative decoding strategies to higher-dimensional systems, including maximum likelihood decoders and decoders that incorporate flag qudit information, to evaluate potential improvements in decoding performance and efficiency. Finally, we hope these results encourage continued experimental efforts in quantum information processing with qudits, as the thresholds obtained in this work support the potential of fault-tolerant quantum computation with logical qudits to reduce both overhead and circuit depth.\\

\appendix

\section{Creating higher dimensional matching graphs}\label{app: A}
In this appendix, we provide a detailed example demonstrating how to construct the matching graph from a higher-dimensional parity check matrix. The amount of rows $R$ in the matrix corresponds to the amount of ancilla qudits, and the entries indicate the powers to which the respective Pauli operators are raised. The matching graph can be created by following the steps:

\begin{enumerate}
    \item Extract the $R$ stabilizers from the parity check matrix.
    \item Create $R(q-1)$ nodes, $q-1$ for every stabilizer. Label them with the stabilizer and one of the $q-1$ possible eigenvalues.
    \item For each labeled node, create dangling edges and label them with the errors that give rise to the corresponding eigenvalue. To determine the errors on the edges surrounding each node, raise the Pauli operator at index $i$ to the power of the corresponding eigenvalue $a$ modulo $q$. Then, interchange $X$ and $Z$ operators, as they measure each other's parity. Finally, take the Hermitian conjugate (dagger), since each error in the syndrome extraction circuit propagates to the ancilla as a $Z$ error and is transformed into the $X$ basis via the conjugation relation given in Eq. \ref{eq: conj}.
    \item Connect nodes that share a dangling edge representing the same error.
    \end{enumerate}

Steps two and three introduce and utilize the expanded set of eigenvalues inherent to qudits, distinguishing this approach from the established method used to construct matching graphs for qubit-based codes, as described in \cite{Antipov_Kiktenko_Fedorov_2023}. Notably, this method is a generalization to qudits of arbitrary prime dimension $q$, and reduces to the qubit case when $q = 2$. To illustrate the procedure, we consider the case $q=3$ using the parity-check matrix presented in the paper:
\[
\left(
\begin{array}{ccccc|ccccc}
  0 & 0 & q-1 & 0 & 1 & 1 & q-1 & 0 & 0 & 0 \\
  1 & 0 & 0 & q-1 & 0 & 0 & 1 & q-1 & 0 & 0 \\
  0 & 1 & 0 & 0 & q-1 & 0 & 0 & 1 & q-1 & 0 \\
  q-1 & 0 & 1 & 0 & 0 & 0 & 0 & 0 & 1 & q-1 \\
\end{array}
\right)
\].

\begin{enumerate}
    \item This parity check matrix has $R=4$ rows. The first row represents ancilla $A_1$ measuring stabilizer $S_1 = XX^{2} Z^{2}IZ$, the second row $A_2$ measuring $S_2 = ZXX^{2} Z^{2}I$, the third row $A_3$ measuring $S_3 = IZXX^{2} Z^{2}$ and the last row $A_4$ measuring $S_4 = Z^{2}IZXX^{2}$.
    \item The 8 labeled nodes can be seen on Figure \ref{fig: make_match_1}. The number above the stabilizer indicates a measurement outcome in the X basis of the corresponding ancilla. It represents the $a$'th eigenvalue $\omega^a$ with  $\omega=e^{\frac{2\pi i}{3}}$. 
    \begin{figure}
        \centering
        \includegraphics[width=0.5\linewidth]{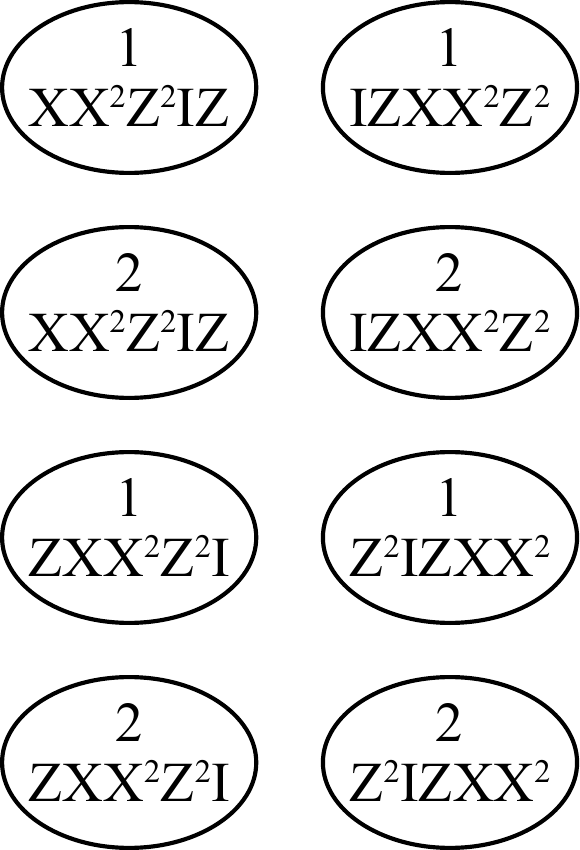}
        \caption{Labeled nodes during the creation of the matching graph for the 5 qutrit code.}
        \label{fig: make_match_1}
    \end{figure}
    \item  The first Pauli operator in the first node is $X$. Raising this operator to the power of $1$, followed by interchanging $X$ and $Z$, and subsequently taking the inverse, results in the error $Z^2$ on the qutrit located at index $0$. We therefore attach a dangling edge labeled by $Z_0^2$ to this node. This process can be repeated for every Pauli operator in the labeled nodes and the result can be seen in figure \ref{fig: make_matchin_2}. It is important to note that, for this particular code, a Pauli operator can only be raised to the power $a$ or $a(q-1) \mod q = -a$. This characteristic contributes to the separation observed in matching graphs for qudit dimensions greater than $3$.
    \begin{figure}
        \centering
        \includegraphics[width=0.8\linewidth]{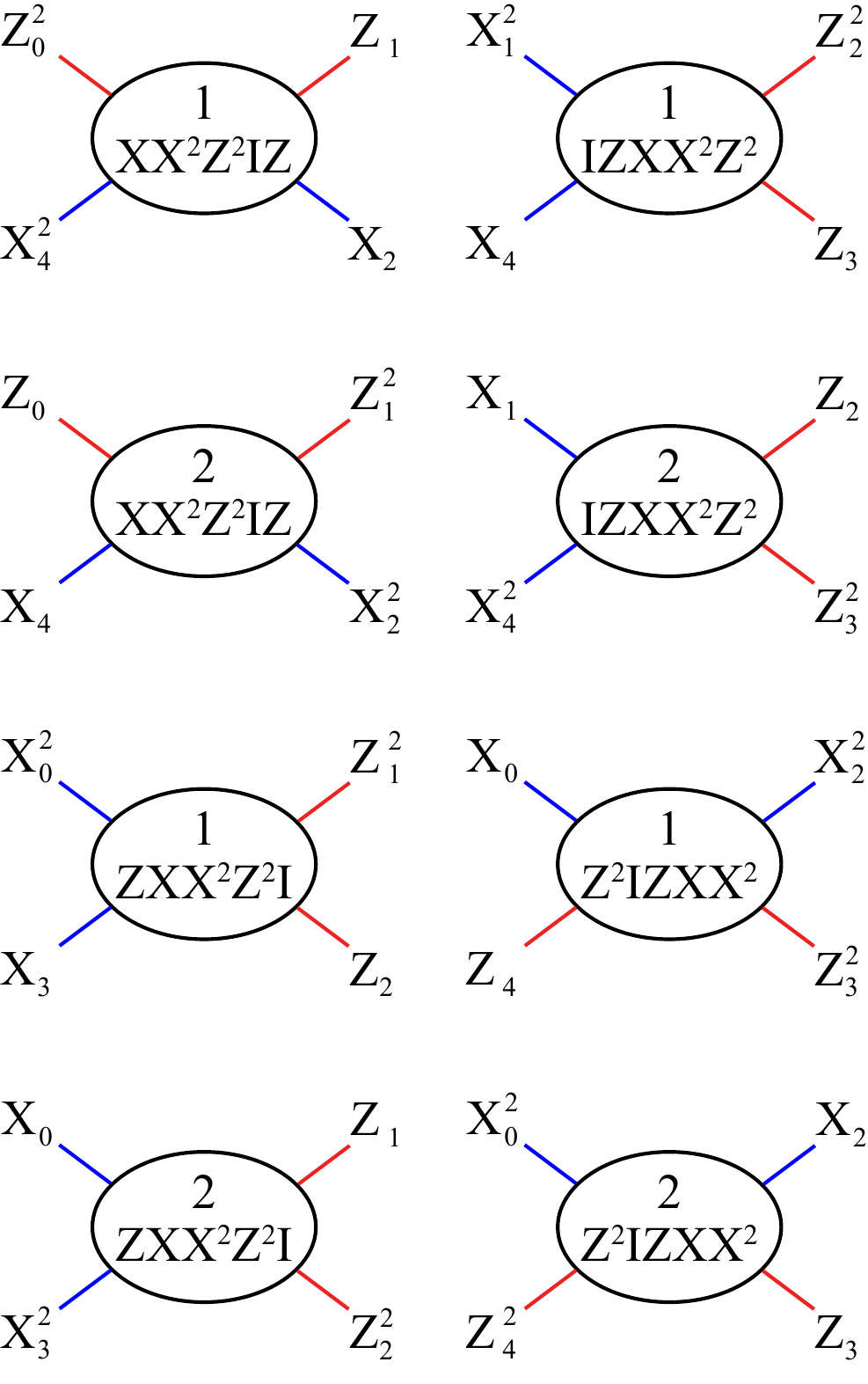}
        \caption{Labeled nodes, with dangling edges representing errors, during the creation of the matching graph for the 5 qutrit code.}
        \label{fig: make_matchin_2}
    \end{figure}
    \item Connecting the nodes that correspond to identical errors and subsequently reordering them yields the matching graph shown in Figure \ref{fig: MG3}. It is worth noting that there are at most two dangling edges per error, as each column of the parity-check matrix contains at most two non-zero entries.
\end{enumerate}
\newpage
\bibliography{ref.bib}

\begin{thebibliography}{45}%
\makeatletter
\providecommand \@ifxundefined [1]{%
 \@ifx{#1\undefined}
}%
\providecommand \@ifnum [1]{%
 \ifnum #1\expandafter \@firstoftwo
 \else \expandafter \@secondoftwo
 \fi
}%
\providecommand \@ifx [1]{%
 \ifx #1\expandafter \@firstoftwo
 \else \expandafter \@secondoftwo
 \fi
}%
\providecommand \natexlab [1]{#1}%
\providecommand \enquote  [1]{``#1''}%
\providecommand \bibnamefont  [1]{#1}%
\providecommand \bibfnamefont [1]{#1}%
\providecommand \citenamefont [1]{#1}%
\providecommand \href@noop [0]{\@secondoftwo}%
\providecommand \href [0]{\begingroup \@sanitize@url \@href}%
\providecommand \@href[1]{\@@startlink{#1}\@@href}%
\providecommand \@@href[1]{\endgroup#1\@@endlink}%
\providecommand \@sanitize@url [0]{\catcode `\\12\catcode `\$12\catcode `\&12\catcode `\#12\catcode `\^12\catcode `\_12\catcode `\%12\relax}%
\providecommand \@@startlink[1]{}%
\providecommand \@@endlink[0]{}%
\providecommand \url  [0]{\begingroup\@sanitize@url \@url }%
\providecommand \@url [1]{\endgroup\@href {#1}{\urlprefix }}%
\providecommand \urlprefix  [0]{URL }%
\providecommand \Eprint [0]{\href }%
\providecommand \doibase [0]{https://doi.org/}%
\providecommand \selectlanguage [0]{\@gobble}%
\providecommand \bibinfo  [0]{\@secondoftwo}%
\providecommand \bibfield  [0]{\@secondoftwo}%
\providecommand \translation [1]{[#1]}%
\providecommand \BibitemOpen [0]{}%
\providecommand \bibitemStop [0]{}%
\providecommand \bibitemNoStop [0]{.\EOS\space}%
\providecommand \EOS [0]{\spacefactor3000\relax}%
\providecommand \BibitemShut  [1]{\csname bibitem#1\endcsname}%
\let\auto@bib@innerbib\@empty
\bibitem [{\citenamefont {Zurek}(1991)}]{Zurek_1991}%
  \BibitemOpen
  \bibfield  {author} {\bibinfo {author} {\bibfnamefont {W.~H.}\ \bibnamefont {Zurek}},\ }\bibfield  {title} {\bibinfo {title} {Decoherence and the transition from quantum to classical},\ }\href {https://doi.org/10.1063/1.881293} {\bibfield  {journal} {\bibinfo  {journal} {Physics Today}\ }\textbf {\bibinfo {volume} {44}},\ \bibinfo {pages} {36–44} (\bibinfo {year} {1991})}\BibitemShut {NoStop}%
\bibitem [{\citenamefont {Acharya}\ \emph {et~al.}(2024)\citenamefont {Acharya}, \citenamefont {et~al.}, \citenamefont {AI},\ and\ \citenamefont {Collaborators}}]{Acharya2024}%
  \BibitemOpen
  \bibfield  {author} {\bibinfo {author} {\bibfnamefont {R.}~\bibnamefont {Acharya}}, \bibinfo {author} {\bibnamefont {et~al.}}, \bibinfo {author} {\bibfnamefont {G.~Q.}\ \bibnamefont {AI}},\ and\ \bibinfo {author} {\bibnamefont {Collaborators}},\ }\bibfield  {title} {\bibinfo {title} {Quantum error correction below the surface code threshold},\ }\href {https://doi.org/10.1038/s41586-024-08449-y} {\bibfield  {journal} {\bibinfo  {journal} {Nature}\ } (\bibinfo {year} {2024})}\BibitemShut {NoStop}%
\bibitem [{\citenamefont {Bravyi}\ \emph {et~al.}(1998)\citenamefont {Bravyi}, \citenamefont {Kitaev},\ and\ \citenamefont {Landau}}]{Bravyi_Kitaev_Landau_1998}%
  \BibitemOpen
  \bibfield  {author} {\bibinfo {author} {\bibfnamefont {S.~B.}\ \bibnamefont {Bravyi}}, \bibinfo {author} {\bibfnamefont {A.~Y.}\ \bibnamefont {Kitaev}},\ and\ \bibinfo {author} {\bibfnamefont {L.~D.}\ \bibnamefont {Landau}},\ }\bibfield  {title} {\bibinfo {title} {Quantum codes on a lattice with boundary},\ }\href@noop {} {\  (\bibinfo {year} {1998})},\ \bibinfo {note} {arXiv:quant-ph/9811052}\BibitemShut {NoStop}%
\bibitem [{\citenamefont {Fowler}\ \emph {et~al.}(2012)\citenamefont {Fowler}, \citenamefont {Mariantoni}, \citenamefont {Martinis},\ and\ \citenamefont {Cleland}}]{Fowler_Mariantoni_Martinis_Cleland_2012}%
  \BibitemOpen
  \bibfield  {author} {\bibinfo {author} {\bibfnamefont {A.~G.}\ \bibnamefont {Fowler}}, \bibinfo {author} {\bibfnamefont {M.}~\bibnamefont {Mariantoni}}, \bibinfo {author} {\bibfnamefont {J.~M.}\ \bibnamefont {Martinis}},\ and\ \bibinfo {author} {\bibfnamefont {A.~N.}\ \bibnamefont {Cleland}},\ }\bibfield  {title} {\bibinfo {title} {Surface codes: Towards practical large-scale quantum computation},\ }\href {https://doi.org/10.1103/PhysRevA.86.032324} {\bibfield  {journal} {\bibinfo  {journal} {Phys. Rev. A}\ }\textbf {\bibinfo {volume} {86}},\ \bibinfo {pages} {032324} (\bibinfo {year} {2012})}\BibitemShut {NoStop}%
\bibitem [{\citenamefont {Panteleev}\ and\ \citenamefont {Kalachev}(2022)}]{Panteleev_Kalachev_2022}%
  \BibitemOpen
  \bibfield  {author} {\bibinfo {author} {\bibfnamefont {P.}~\bibnamefont {Panteleev}}\ and\ \bibinfo {author} {\bibfnamefont {G.}~\bibnamefont {Kalachev}},\ }\bibfield  {title} {\bibinfo {title} {Asymptotically good quantum and locally testable classical ldpc codes},\ }in\ \href {https://doi.org/10.1145/3519935.3520017} {\emph {\bibinfo {booktitle} {Proceedings of the 54th Annual ACM SIGACT Symposium on Theory of Computing}}},\ \bibinfo {series and number} {STOC 2022}\ (\bibinfo  {publisher} {Association for Computing Machinery},\ \bibinfo {address} {New York, NY, USA},\ \bibinfo {year} {2022})\ p.\ \bibinfo {pages} {375–388}\BibitemShut {NoStop}%
\bibitem [{\citenamefont {Bravyi}\ \emph {et~al.}(2024)\citenamefont {Bravyi}, \citenamefont {Cross}, \citenamefont {Gambetta}, \citenamefont {Maslov}, \citenamefont {Rall},\ and\ \citenamefont {Yoder}}]{Bravyi_Cross_Gambetta_Maslov_Rall_Yoder_2024}%
  \BibitemOpen
  \bibfield  {author} {\bibinfo {author} {\bibfnamefont {S.}~\bibnamefont {Bravyi}}, \bibinfo {author} {\bibfnamefont {A.~W.}\ \bibnamefont {Cross}}, \bibinfo {author} {\bibfnamefont {J.~M.}\ \bibnamefont {Gambetta}}, \bibinfo {author} {\bibfnamefont {D.}~\bibnamefont {Maslov}}, \bibinfo {author} {\bibfnamefont {P.}~\bibnamefont {Rall}},\ and\ \bibinfo {author} {\bibfnamefont {T.~J.}\ \bibnamefont {Yoder}},\ }\bibfield  {title} {\bibinfo {title} {High-threshold and low-overhead fault-tolerant quantum memory},\ }\href {https://doi.org/10.1038/s41586-024-07107-7} {\bibfield  {journal} {\bibinfo  {journal} {Nature}\ }\textbf {\bibinfo {volume} {627}},\ \bibinfo {pages} {778–782} (\bibinfo {year} {2024})}\BibitemShut {NoStop}%
\bibitem [{\citenamefont {Field}\ and\ \citenamefont {Simula}(2018)}]{Field_Simula_2018}%
  \BibitemOpen
  \bibfield  {author} {\bibinfo {author} {\bibfnamefont {B.}~\bibnamefont {Field}}\ and\ \bibinfo {author} {\bibfnamefont {T.}~\bibnamefont {Simula}},\ }\bibfield  {title} {\bibinfo {title} {Introduction to topological quantum computation with non-abelian anyons},\ }\href {https://doi.org/10.1088/2058-9565/aacad2} {\bibfield  {journal} {\bibinfo  {journal} {Quantum Science and Technology}\ }\textbf {\bibinfo {volume} {3}},\ \bibinfo {pages} {045004} (\bibinfo {year} {2018})}\BibitemShut {NoStop}%
\bibitem [{\citenamefont {Schotte}\ \emph {et~al.}(2022)\citenamefont {Schotte}, \citenamefont {Zhu}, \citenamefont {Burgelman},\ and\ \citenamefont {Verstraete}}]{Schotte_Zhu_Burgelman_Verstraete_2022}%
  \BibitemOpen
  \bibfield  {author} {\bibinfo {author} {\bibfnamefont {A.}~\bibnamefont {Schotte}}, \bibinfo {author} {\bibfnamefont {G.}~\bibnamefont {Zhu}}, \bibinfo {author} {\bibfnamefont {L.}~\bibnamefont {Burgelman}},\ and\ \bibinfo {author} {\bibfnamefont {F.}~\bibnamefont {Verstraete}},\ }\bibfield  {title} {\bibinfo {title} {Quantum error correction thresholds for the universal fibonacci turaev-viro code},\ }\href {https://doi.org/10.1103/PhysRevX.12.021012} {\bibfield  {journal} {\bibinfo  {journal} {Physical Review X}\ }\textbf {\bibinfo {volume} {12}},\ \bibinfo {pages} {021012} (\bibinfo {year} {2022})}\BibitemShut {NoStop}%
\bibitem [{\citenamefont {Wang}\ \emph {et~al.}(2020)\citenamefont {Wang}, \citenamefont {Hu}, \citenamefont {Sanders},\ and\ \citenamefont {Kais}}]{Wang_Hu_Sanders_Kais_2020}%
  \BibitemOpen
  \bibfield  {author} {\bibinfo {author} {\bibfnamefont {Y.}~\bibnamefont {Wang}}, \bibinfo {author} {\bibfnamefont {Z.}~\bibnamefont {Hu}}, \bibinfo {author} {\bibfnamefont {B.~C.}\ \bibnamefont {Sanders}},\ and\ \bibinfo {author} {\bibfnamefont {S.}~\bibnamefont {Kais}},\ }\bibfield  {title} {\bibinfo {title} {Qudits and high-dimensional quantum computing},\ }\href@noop {} {\bibfield  {journal} {\bibinfo  {journal} {Frontiers in Physics}\ }\textbf {\bibinfo {volume} {8}} (\bibinfo {year} {2020})}\BibitemShut {NoStop}%
\bibitem [{\citenamefont {Huber}\ and\ \citenamefont {de~Vicente}(2013)}]{Huber_de_Vicente_2013}%
  \BibitemOpen
  \bibfield  {author} {\bibinfo {author} {\bibfnamefont {M.}~\bibnamefont {Huber}}\ and\ \bibinfo {author} {\bibfnamefont {J.~I.}\ \bibnamefont {de~Vicente}},\ }\bibfield  {title} {\bibinfo {title} {Structure of multidimensional entanglement in multipartite systems},\ }\href {https://doi.org/10.1103/PhysRevLett.110.030501} {\bibfield  {journal} {\bibinfo  {journal} {Physical Review Letters}\ }\textbf {\bibinfo {volume} {110}},\ \bibinfo {pages} {030501} (\bibinfo {year} {2013})}\BibitemShut {NoStop}%
\bibitem [{\citenamefont {Gokhale}\ \emph {et~al.}(2019)\citenamefont {Gokhale}, \citenamefont {Baker}, \citenamefont {Duckering}, \citenamefont {Brown}, \citenamefont {Brown},\ and\ \citenamefont {Chong}}]{Gokhale_Baker_Duckering_Brown_Brown_Chong_2019}%
  \BibitemOpen
  \bibfield  {author} {\bibinfo {author} {\bibfnamefont {P.}~\bibnamefont {Gokhale}}, \bibinfo {author} {\bibfnamefont {J.~M.}\ \bibnamefont {Baker}}, \bibinfo {author} {\bibfnamefont {C.}~\bibnamefont {Duckering}}, \bibinfo {author} {\bibfnamefont {N.~C.}\ \bibnamefont {Brown}}, \bibinfo {author} {\bibfnamefont {K.~R.}\ \bibnamefont {Brown}},\ and\ \bibinfo {author} {\bibfnamefont {F.~T.}\ \bibnamefont {Chong}},\ }\bibfield  {title} {\bibinfo {title} {Asymptotic improvements to quantum circuits via qutrits},\ }in\ \href {https://doi.org/10.1145/3307650.3322253} {\emph {\bibinfo {booktitle} {Proceedings of the 46th International Symposium on Computer Architecture}}}\ (\bibinfo  {publisher} {ACM},\ \bibinfo {address} {Phoenix Arizona},\ \bibinfo {year} {2019})\ p.\ \bibinfo {pages} {554–566}\BibitemShut {NoStop}%
\bibitem [{\citenamefont {Campbell}\ \emph {et~al.}(2012)\citenamefont {Campbell}, \citenamefont {Anwar},\ and\ \citenamefont {Browne}}]{Campbell_Anwar_Browne_2012}%
  \BibitemOpen
  \bibfield  {author} {\bibinfo {author} {\bibfnamefont {E.~T.}\ \bibnamefont {Campbell}}, \bibinfo {author} {\bibfnamefont {H.}~\bibnamefont {Anwar}},\ and\ \bibinfo {author} {\bibfnamefont {D.~E.}\ \bibnamefont {Browne}},\ }\bibfield  {title} {\bibinfo {title} {Magic-state distillation in all prime dimensions using quantum reed-muller codes},\ }\href {https://doi.org/10.1103/PhysRevX.2.041021} {\bibfield  {journal} {\bibinfo  {journal} {Physical Review X}\ }\textbf {\bibinfo {volume} {2}},\ \bibinfo {pages} {041021} (\bibinfo {year} {2012})}\BibitemShut {NoStop}%
\bibitem [{\citenamefont {Campbell}(2014)}]{Campbell_2014}%
  \BibitemOpen
  \bibfield  {author} {\bibinfo {author} {\bibfnamefont {E.~T.}\ \bibnamefont {Campbell}},\ }\bibfield  {title} {\bibinfo {title} {Enhanced fault-tolerant quantum computing in d -level systems},\ }\href {https://doi.org/10.1103/PhysRevLett.113.230501} {\bibfield  {journal} {\bibinfo  {journal} {Physical Review Letters}\ }\textbf {\bibinfo {volume} {113}},\ \bibinfo {pages} {230501} (\bibinfo {year} {2014})}\BibitemShut {NoStop}%
\bibitem [{\citenamefont {Chi}\ \emph {et~al.}(2022)\citenamefont {Chi}, \citenamefont {Huang}, \citenamefont {Zhang}, \citenamefont {Mao}, \citenamefont {Zhou}, \citenamefont {Chen}, \citenamefont {Zhai}, \citenamefont {Bao}, \citenamefont {Dai}, \citenamefont {Yuan}, \citenamefont {Zhang}, \citenamefont {Dai}, \citenamefont {Tang}, \citenamefont {Yang}, \citenamefont {Li}, \citenamefont {Ding}, \citenamefont {Oxenløwe}, \citenamefont {Thompson}, \citenamefont {O'Brien}, \citenamefont {Li}, \citenamefont {Gong},\ and\ \citenamefont {Wang}}]{Chi_etal._2022}%
  \BibitemOpen
  \bibfield  {author} {\bibinfo {author} {\bibfnamefont {Y.}~\bibnamefont {Chi}}, \bibinfo {author} {\bibfnamefont {J.}~\bibnamefont {Huang}}, \bibinfo {author} {\bibfnamefont {Z.}~\bibnamefont {Zhang}}, \bibinfo {author} {\bibfnamefont {J.}~\bibnamefont {Mao}}, \bibinfo {author} {\bibfnamefont {Z.}~\bibnamefont {Zhou}}, \bibinfo {author} {\bibfnamefont {X.}~\bibnamefont {Chen}}, \bibinfo {author} {\bibfnamefont {C.}~\bibnamefont {Zhai}}, \bibinfo {author} {\bibfnamefont {J.}~\bibnamefont {Bao}}, \bibinfo {author} {\bibfnamefont {T.}~\bibnamefont {Dai}}, \bibinfo {author} {\bibfnamefont {H.}~\bibnamefont {Yuan}}, \bibinfo {author} {\bibfnamefont {M.}~\bibnamefont {Zhang}}, \bibinfo {author} {\bibfnamefont {D.}~\bibnamefont {Dai}}, \bibinfo {author} {\bibfnamefont {B.}~\bibnamefont {Tang}}, \bibinfo {author} {\bibfnamefont {Y.}~\bibnamefont {Yang}}, \bibinfo {author} {\bibfnamefont {Z.}~\bibnamefont {Li}}, \bibinfo {author} {\bibfnamefont {Y.}~\bibnamefont {Ding}}, \bibinfo {author} {\bibfnamefont {L.~K.}\
  \bibnamefont {Oxenløwe}}, \bibinfo {author} {\bibfnamefont {M.~G.}\ \bibnamefont {Thompson}}, \bibinfo {author} {\bibfnamefont {J.~L.}\ \bibnamefont {O'Brien}}, \bibinfo {author} {\bibfnamefont {Y.}~\bibnamefont {Li}}, \bibinfo {author} {\bibfnamefont {Q.}~\bibnamefont {Gong}},\ and\ \bibinfo {author} {\bibfnamefont {J.}~\bibnamefont {Wang}},\ }\bibfield  {title} {\bibinfo {title} {A programmable qudit-based quantum processor},\ }\href@noop {} {\bibfield  {journal} {\bibinfo  {journal} {Nature Communications}\ }\textbf {\bibinfo {volume} {13}} (\bibinfo {year} {2022})}\BibitemShut {NoStop}%
\bibitem [{\citenamefont {Lanyon}\ \emph {et~al.}(2009)\citenamefont {Lanyon}, \citenamefont {Barbieri}, \citenamefont {Almeida}, \citenamefont {Jennewein}, \citenamefont {Ralph}, \citenamefont {Resch}, \citenamefont {Pryde}, \citenamefont {O'Brien}, \citenamefont {Gilchrist},\ and\ \citenamefont {White}}]{Lanyon_Barbieri_Almeida_Jennewein_Ralph_Resch_Pryde_O'Brien_Gilchrist_White_2009}%
  \BibitemOpen
  \bibfield  {author} {\bibinfo {author} {\bibfnamefont {B.~P.}\ \bibnamefont {Lanyon}}, \bibinfo {author} {\bibfnamefont {M.}~\bibnamefont {Barbieri}}, \bibinfo {author} {\bibfnamefont {M.~P.}\ \bibnamefont {Almeida}}, \bibinfo {author} {\bibfnamefont {T.}~\bibnamefont {Jennewein}}, \bibinfo {author} {\bibfnamefont {T.~C.}\ \bibnamefont {Ralph}}, \bibinfo {author} {\bibfnamefont {K.~J.}\ \bibnamefont {Resch}}, \bibinfo {author} {\bibfnamefont {G.~J.}\ \bibnamefont {Pryde}}, \bibinfo {author} {\bibfnamefont {J.~L.}\ \bibnamefont {O'Brien}}, \bibinfo {author} {\bibfnamefont {A.}~\bibnamefont {Gilchrist}},\ and\ \bibinfo {author} {\bibfnamefont {A.~G.}\ \bibnamefont {White}},\ }\bibfield  {title} {\bibinfo {title} {Simplifying quantum logic using higher-dimensional hilbert spaces},\ }\href {https://doi.org/10.1038/nphys1150} {\bibfield  {journal} {\bibinfo  {journal} {Nature Physics}\ }\textbf {\bibinfo {volume} {5}},\ \bibinfo {pages} {134–140} (\bibinfo {year} {2009})}\BibitemShut {NoStop}%
\bibitem [{\citenamefont {Bianchetti}\ \emph {et~al.}(2010)\citenamefont {Bianchetti}, \citenamefont {Filipp}, \citenamefont {Baur}, \citenamefont {Fink}, \citenamefont {Lang}, \citenamefont {Steffen}, \citenamefont {Boissonneault}, \citenamefont {Blais},\ and\ \citenamefont {Wallraff}}]{Bianchetti_Filipp_Baur_Fink_Lang_Steffen_Boissonneault_Blais_Wallraff_2010}%
  \BibitemOpen
  \bibfield  {author} {\bibinfo {author} {\bibfnamefont {R.}~\bibnamefont {Bianchetti}}, \bibinfo {author} {\bibfnamefont {S.}~\bibnamefont {Filipp}}, \bibinfo {author} {\bibfnamefont {M.}~\bibnamefont {Baur}}, \bibinfo {author} {\bibfnamefont {J.~M.}\ \bibnamefont {Fink}}, \bibinfo {author} {\bibfnamefont {C.}~\bibnamefont {Lang}}, \bibinfo {author} {\bibfnamefont {L.}~\bibnamefont {Steffen}}, \bibinfo {author} {\bibfnamefont {M.}~\bibnamefont {Boissonneault}}, \bibinfo {author} {\bibfnamefont {A.}~\bibnamefont {Blais}},\ and\ \bibinfo {author} {\bibfnamefont {A.}~\bibnamefont {Wallraff}},\ }\bibfield  {title} {\bibinfo {title} {Control and tomography of a three level superconducting artificial atom},\ }\href {https://doi.org/10.1103/PhysRevLett.105.223601} {\bibfield  {journal} {\bibinfo  {journal} {Physical Review Letters}\ }\textbf {\bibinfo {volume} {105}},\ \bibinfo {pages} {223601} (\bibinfo {year} {2010})}\BibitemShut {NoStop}%
\bibitem [{\citenamefont {Kononenko}\ \emph {et~al.}(2021)\citenamefont {Kononenko}, \citenamefont {Yurtalan}, \citenamefont {Ren}, \citenamefont {Shi}, \citenamefont {Ashhab},\ and\ \citenamefont {Lupascu}}]{Kononenko_Yurtalan_Ren_Shi_Ashhab_Lupascu_2021}%
  \BibitemOpen
  \bibfield  {author} {\bibinfo {author} {\bibfnamefont {M.}~\bibnamefont {Kononenko}}, \bibinfo {author} {\bibfnamefont {M.~A.}\ \bibnamefont {Yurtalan}}, \bibinfo {author} {\bibfnamefont {S.}~\bibnamefont {Ren}}, \bibinfo {author} {\bibfnamefont {J.}~\bibnamefont {Shi}}, \bibinfo {author} {\bibfnamefont {S.}~\bibnamefont {Ashhab}},\ and\ \bibinfo {author} {\bibfnamefont {A.}~\bibnamefont {Lupascu}},\ }\bibfield  {title} {\bibinfo {title} {Characterization of control in a superconducting qutrit using randomized benchmarking},\ }\href {https://doi.org/10.1103/PhysRevResearch.3.L042007} {\bibfield  {journal} {\bibinfo  {journal} {Physical Review Research}\ }\textbf {\bibinfo {volume} {3}},\ \bibinfo {pages} {L042007} (\bibinfo {year} {2021})}\BibitemShut {NoStop}%
\bibitem [{\citenamefont {Fernández~de Fuentes}\ \emph {et~al.}(2024)\citenamefont {Fernández~de Fuentes}, \citenamefont {Botzem}, \citenamefont {Johnson}, \citenamefont {Vaartjes}, \citenamefont {Asaad}, \citenamefont {Mourik}, \citenamefont {Hudson}, \citenamefont {Itoh}, \citenamefont {Johnson}, \citenamefont {Jakob}, \citenamefont {McCallum}, \citenamefont {Jamieson}, \citenamefont {Dzurak},\ and\ \citenamefont {Morello}}]{deFuentes_etal._2023}%
  \BibitemOpen
  \bibfield  {author} {\bibinfo {author} {\bibfnamefont {I.}~\bibnamefont {Fernández~de Fuentes}}, \bibinfo {author} {\bibfnamefont {T.}~\bibnamefont {Botzem}}, \bibinfo {author} {\bibfnamefont {M.~A.~I.}\ \bibnamefont {Johnson}}, \bibinfo {author} {\bibfnamefont {A.}~\bibnamefont {Vaartjes}}, \bibinfo {author} {\bibfnamefont {S.}~\bibnamefont {Asaad}}, \bibinfo {author} {\bibfnamefont {V.}~\bibnamefont {Mourik}}, \bibinfo {author} {\bibfnamefont {F.~E.}\ \bibnamefont {Hudson}}, \bibinfo {author} {\bibfnamefont {K.~M.}\ \bibnamefont {Itoh}}, \bibinfo {author} {\bibfnamefont {B.~C.}\ \bibnamefont {Johnson}}, \bibinfo {author} {\bibfnamefont {A.~M.}\ \bibnamefont {Jakob}}, \bibinfo {author} {\bibfnamefont {J.~C.}\ \bibnamefont {McCallum}}, \bibinfo {author} {\bibfnamefont {D.~N.}\ \bibnamefont {Jamieson}}, \bibinfo {author} {\bibfnamefont {A.~S.}\ \bibnamefont {Dzurak}},\ and\ \bibinfo {author} {\bibfnamefont {A.}~\bibnamefont {Morello}},\ }\bibfield  {title} {\bibinfo {title} {Navigating the 16-dimensional
  hilbert space of a high-spin donor qudit with electric and magnetic fields},\ }\href {https://doi.org/10.1038/s41467-024-45368-y} {\bibfield  {journal} {\bibinfo  {journal} {Nature Communications}\ }\textbf {\bibinfo {volume} {15}},\ \bibinfo {pages} {1380} (\bibinfo {year} {2024})}\BibitemShut {NoStop}%
\bibitem [{\citenamefont {Godfrin}\ \emph {et~al.}(2017)\citenamefont {Godfrin}, \citenamefont {Ferhat}, \citenamefont {Ballou}, \citenamefont {Klyatskaya}, \citenamefont {Ruben}, \citenamefont {Wernsdorfer},\ and\ \citenamefont {Balestro}}]{Godfrin_Ferhat_Ballou_Klyatskaya_Ruben_Wernsdorfer_Balestro_2017}%
  \BibitemOpen
  \bibfield  {author} {\bibinfo {author} {\bibfnamefont {C.}~\bibnamefont {Godfrin}}, \bibinfo {author} {\bibfnamefont {A.}~\bibnamefont {Ferhat}}, \bibinfo {author} {\bibfnamefont {R.}~\bibnamefont {Ballou}}, \bibinfo {author} {\bibfnamefont {S.}~\bibnamefont {Klyatskaya}}, \bibinfo {author} {\bibfnamefont {M.}~\bibnamefont {Ruben}}, \bibinfo {author} {\bibfnamefont {W.}~\bibnamefont {Wernsdorfer}},\ and\ \bibinfo {author} {\bibfnamefont {F.}~\bibnamefont {Balestro}},\ }\bibfield  {title} {\bibinfo {title} {Operating quantum states in single magnetic molecules: Implementation of grover's quantum algorithm},\ }\href {https://doi.org/10.1103/PhysRevLett.119.187702} {\bibfield  {journal} {\bibinfo  {journal} {Physical Review Letters}\ }\textbf {\bibinfo {volume} {119}},\ \bibinfo {pages} {187702} (\bibinfo {year} {2017})}\BibitemShut {NoStop}%
\bibitem [{\citenamefont {Ringbauer}\ \emph {et~al.}(2022)\citenamefont {Ringbauer}, \citenamefont {Meth}, \citenamefont {Postler}, \citenamefont {Stricker}, \citenamefont {Blatt}, \citenamefont {Schindler},\ and\ \citenamefont {Monz}}]{Ringbauer_Meth_Postler_Stricker_Blatt_Schindler_Monz_2022}%
  \BibitemOpen
  \bibfield  {author} {\bibinfo {author} {\bibfnamefont {M.}~\bibnamefont {Ringbauer}}, \bibinfo {author} {\bibfnamefont {M.}~\bibnamefont {Meth}}, \bibinfo {author} {\bibfnamefont {L.}~\bibnamefont {Postler}}, \bibinfo {author} {\bibfnamefont {R.}~\bibnamefont {Stricker}}, \bibinfo {author} {\bibfnamefont {R.}~\bibnamefont {Blatt}}, \bibinfo {author} {\bibfnamefont {P.}~\bibnamefont {Schindler}},\ and\ \bibinfo {author} {\bibfnamefont {T.}~\bibnamefont {Monz}},\ }\bibfield  {title} {\bibinfo {title} {A universal qudit quantum processor with trapped ions},\ }\href {https://doi.org/10.1038/s41567-022-01658-0} {\bibfield  {journal} {\bibinfo  {journal} {Nature Physics}\ }\textbf {\bibinfo {volume} {18}},\ \bibinfo {pages} {1053–1057} (\bibinfo {year} {2022})}\BibitemShut {NoStop}%
\bibitem [{\citenamefont {Weggemans}\ \emph {et~al.}(2022)\citenamefont {Weggemans}, \citenamefont {Urech}, \citenamefont {Rausch}, \citenamefont {Spreeuw}, \citenamefont {Boucherie}, \citenamefont {Schreck}, \citenamefont {Schoutens}, \citenamefont {Minář},\ and\ \citenamefont {Speelman}}]{Weggemans_Urech_Rausch_Spreeuw_Boucherie_Schreck_Schoutens_Minář_Speelman_2022}%
  \BibitemOpen
  \bibfield  {author} {\bibinfo {author} {\bibfnamefont {J.~R.}\ \bibnamefont {Weggemans}}, \bibinfo {author} {\bibfnamefont {A.}~\bibnamefont {Urech}}, \bibinfo {author} {\bibfnamefont {A.}~\bibnamefont {Rausch}}, \bibinfo {author} {\bibfnamefont {R.}~\bibnamefont {Spreeuw}}, \bibinfo {author} {\bibfnamefont {R.}~\bibnamefont {Boucherie}}, \bibinfo {author} {\bibfnamefont {F.}~\bibnamefont {Schreck}}, \bibinfo {author} {\bibfnamefont {K.}~\bibnamefont {Schoutens}}, \bibinfo {author} {\bibfnamefont {J.}~\bibnamefont {Minář}},\ and\ \bibinfo {author} {\bibfnamefont {F.}~\bibnamefont {Speelman}},\ }\bibfield  {title} {\bibinfo {title} {Solving correlation clustering with qaoa and a rydberg qudit system: a full-stack approach},\ }\href {https://doi.org/10.22331/q-2022-04-13-687} {\bibfield  {journal} {\bibinfo  {journal} {Quantum}\ }\textbf {\bibinfo {volume} {6}},\ \bibinfo {pages} {687} (\bibinfo {year} {2022})}\BibitemShut {NoStop}%
\bibitem [{\citenamefont {Ahn}\ \emph {et~al.}(2000)\citenamefont {Ahn}, \citenamefont {Weinacht},\ and\ \citenamefont {Bucksbaum}}]{Ahn_Weinacht_Bucksbaum_2000}%
  \BibitemOpen
  \bibfield  {author} {\bibinfo {author} {\bibfnamefont {J.}~\bibnamefont {Ahn}}, \bibinfo {author} {\bibfnamefont {T.~C.}\ \bibnamefont {Weinacht}},\ and\ \bibinfo {author} {\bibfnamefont {P.~H.}\ \bibnamefont {Bucksbaum}},\ }\bibfield  {title} {\bibinfo {title} {Information storage and retrieval through quantum phase},\ }\href {https://doi.org/10.1126/science.287.5452.463} {\bibfield  {journal} {\bibinfo  {journal} {Science}\ }\textbf {\bibinfo {volume} {287}},\ \bibinfo {pages} {463–465} (\bibinfo {year} {2000})}\BibitemShut {NoStop}%
\bibitem [{\citenamefont {Laflamme}\ \emph {et~al.}(1996)\citenamefont {Laflamme}, \citenamefont {Miquel}, \citenamefont {Paz},\ and\ \citenamefont {Zurek}}]{Laflamme_Miquel_Paz_Zurek_1996}%
  \BibitemOpen
  \bibfield  {author} {\bibinfo {author} {\bibfnamefont {R.}~\bibnamefont {Laflamme}}, \bibinfo {author} {\bibfnamefont {C.}~\bibnamefont {Miquel}}, \bibinfo {author} {\bibfnamefont {J.~P.}\ \bibnamefont {Paz}},\ and\ \bibinfo {author} {\bibfnamefont {W.~H.}\ \bibnamefont {Zurek}},\ }\bibfield  {title} {\bibinfo {title} {Perfect quantum error correcting code},\ }\href {https://doi.org/10.1103/PhysRevLett.77.198} {\bibfield  {journal} {\bibinfo  {journal} {Phys. Rev. Lett.}\ }\textbf {\bibinfo {volume} {77}},\ \bibinfo {pages} {198} (\bibinfo {year} {1996})}\BibitemShut {NoStop}%
\bibitem [{\citenamefont {Muthukrishnan}\ and\ \citenamefont {Stroud}(2000)}]{Muthukrishnan_Stroud_2000}%
  \BibitemOpen
  \bibfield  {author} {\bibinfo {author} {\bibfnamefont {A.}~\bibnamefont {Muthukrishnan}}\ and\ \bibinfo {author} {\bibfnamefont {C.~R.}\ \bibnamefont {Stroud}},\ }\bibfield  {title} {\bibinfo {title} {Multivalued logic gates for quantum computation},\ }\href {https://doi.org/10.1103/PhysRevA.62.052309} {\bibfield  {journal} {\bibinfo  {journal} {Physical Review A}\ }\textbf {\bibinfo {volume} {62}},\ \bibinfo {pages} {052309} (\bibinfo {year} {2000})}\BibitemShut {NoStop}%
\bibitem [{\citenamefont {Brennen}\ \emph {et~al.}(2005)\citenamefont {Brennen}, \citenamefont {O'Leary},\ and\ \citenamefont {Bullock}}]{Brennen_O'Leary_Bullock_2005}%
  \BibitemOpen
  \bibfield  {author} {\bibinfo {author} {\bibfnamefont {G.}~\bibnamefont {Brennen}}, \bibinfo {author} {\bibfnamefont {D.}~\bibnamefont {O'Leary}},\ and\ \bibinfo {author} {\bibfnamefont {S.}~\bibnamefont {Bullock}},\ }\bibfield  {title} {\bibinfo {title} {Criteria for exact qudit universality},\ }\href {https://doi.org/10.1103/PhysRevA.71.052318} {\bibfield  {journal} {\bibinfo  {journal} {Physical Review A}\ }\textbf {\bibinfo {volume} {71}},\ \bibinfo {pages} {052318} (\bibinfo {year} {2005})}\BibitemShut {NoStop}%
\bibitem [{\citenamefont {Luo}\ and\ \citenamefont {Wang}(2014)}]{Luo_Wang_2014}%
  \BibitemOpen
  \bibfield  {author} {\bibinfo {author} {\bibfnamefont {M.}~\bibnamefont {Luo}}\ and\ \bibinfo {author} {\bibfnamefont {X.}~\bibnamefont {Wang}},\ }\bibfield  {title} {\bibinfo {title} {Universal quantum computation with qudits},\ }\href {https://doi.org/10.1007/s11433-014-5551-9} {\bibfield  {journal} {\bibinfo  {journal} {Science China: Physics, Mechanics and Astronomy}\ }\textbf {\bibinfo {volume} {57}},\ \bibinfo {pages} {1712–1717} (\bibinfo {year} {2014})}\BibitemShut {NoStop}%
\bibitem [{\citenamefont {Moussa}(2016)}]{Moussa_2016}%
  \BibitemOpen
  \bibfield  {author} {\bibinfo {author} {\bibfnamefont {J.~E.}\ \bibnamefont {Moussa}},\ }\bibfield  {title} {\bibinfo {title} {Transversal clifford gates on folded surface codes},\ }\href {https://doi.org/10.1103/PhysRevA.94.042316} {\bibfield  {journal} {\bibinfo  {journal} {Physical Review A}\ }\textbf {\bibinfo {volume} {94}},\ \bibinfo {pages} {042316} (\bibinfo {year} {2016})}\BibitemShut {NoStop}%
\bibitem [{\citenamefont {Gottesman}(1999)}]{Gottesman_1998}%
  \BibitemOpen
  \bibfield  {author} {\bibinfo {author} {\bibfnamefont {D.}~\bibnamefont {Gottesman}},\ }\bibfield  {title} {\bibinfo {title} {Fault-tolerant quantum computation with higher-dimensional systems},\ }in\ \href@noop {} {\emph {\bibinfo {booktitle} {Quantum Computing and Quantum Communications}}},\ \bibinfo {editor} {edited by\ \bibinfo {editor} {\bibfnamefont {C.~P.}\ \bibnamefont {Williams}}}\ (\bibinfo  {publisher} {Springer Berlin Heidelberg},\ \bibinfo {address} {Berlin, Heidelberg},\ \bibinfo {year} {1999})\ pp.\ \bibinfo {pages} {302--313}\BibitemShut {NoStop}%
\bibitem [{\citenamefont {Gottesman}(1997)}]{Gottesman_1997}%
  \BibitemOpen
  \bibfield  {author} {\bibinfo {author} {\bibfnamefont {D.}~\bibnamefont {Gottesman}},\ }\emph {\bibinfo {title} {Stabilizer Codes and Quantum Error Correction}},\ \href@noop {} {Ph.D. thesis},\ \bibinfo  {school} {California Institute of Technology} (\bibinfo {year} {1997}),\ \bibinfo {note} {arXiv:quant-ph/9705052}\BibitemShut {NoStop}%
\bibitem [{\citenamefont {Gottesman}(1998)}]{Gottesman_qubit}%
  \BibitemOpen
  \bibfield  {author} {\bibinfo {author} {\bibfnamefont {D.}~\bibnamefont {Gottesman}},\ }\bibfield  {title} {\bibinfo {title} {Theory of fault-tolerant quantum computation},\ }\href {https://doi.org/10.1103/PhysRevA.57.127} {\bibfield  {journal} {\bibinfo  {journal} {Physical Review A}\ }\textbf {\bibinfo {volume} {57}},\ \bibinfo {pages} {127–137} (\bibinfo {year} {1998})}\BibitemShut {NoStop}%
\bibitem [{\citenamefont {Knill}\ and\ \citenamefont {Laflamme}(1997)}]{Knill_Laflamme_1997}%
  \BibitemOpen
  \bibfield  {author} {\bibinfo {author} {\bibfnamefont {E.}~\bibnamefont {Knill}}\ and\ \bibinfo {author} {\bibfnamefont {R.}~\bibnamefont {Laflamme}},\ }\bibfield  {title} {\bibinfo {title} {Theory of quantum error-correcting codes},\ }\href {https://doi.org/10.1103/PhysRevA.55.900} {\bibfield  {journal} {\bibinfo  {journal} {Phys. Rev. A}\ }\textbf {\bibinfo {volume} {55}},\ \bibinfo {pages} {900} (\bibinfo {year} {1997})}\BibitemShut {NoStop}%
\bibitem [{\citenamefont {Grassl}\ \emph {et~al.}(2002)\citenamefont {Grassl}, \citenamefont {Otteler},\ and\ \citenamefont {Beth}}]{Grassl_Otteler_Beth_2002}%
  \BibitemOpen
  \bibfield  {author} {\bibinfo {author} {\bibfnamefont {M.}~\bibnamefont {Grassl}}, \bibinfo {author} {\bibfnamefont {M.~R.~Â.}\ \bibnamefont {Otteler}},\ and\ \bibinfo {author} {\bibfnamefont {T.}~\bibnamefont {Beth}},\ }\bibfield  {title} {\bibinfo {title} {Efficient quantum circuits for non-qubit quantum error-correcting codes},\ }\href@noop {} {\bibfield  {journal} {\bibinfo  {journal} {International Journal of Foundations of Computer Science}\ } (\bibinfo {year} {2002})}\BibitemShut {NoStop}%
\bibitem [{\citenamefont {Cross}\ \emph {et~al.}(2009)\citenamefont {Cross}, \citenamefont {DiVincenzo},\ and\ \citenamefont {Terhal}}]{Cross_DiVincenzo_Terhal_2009}%
  \BibitemOpen
  \bibfield  {author} {\bibinfo {author} {\bibfnamefont {A.~W.}\ \bibnamefont {Cross}}, \bibinfo {author} {\bibfnamefont {D.~P.}\ \bibnamefont {DiVincenzo}},\ and\ \bibinfo {author} {\bibfnamefont {B.~M.}\ \bibnamefont {Terhal}},\ }\href {https://doi.org/10.48550/arXiv.0711.1556} {\bibinfo {title} {A comparative code study for quantum fault-tolerance}} (\bibinfo {year} {2009}),\ \bibinfo {note} {arXiv:0711.1556 [quant-ph]}\BibitemShut {NoStop}%
\bibitem [{\citenamefont {Higgott}(2022)}]{Higgott_2022}%
  \BibitemOpen
  \bibfield  {author} {\bibinfo {author} {\bibfnamefont {O.}~\bibnamefont {Higgott}},\ }\bibfield  {title} {\bibinfo {title} {Pymatching: A python package for decoding quantum codes with minimum-weight perfect matching},\ }\href {https://doi.org/10.1145/3505637} {\bibfield  {journal} {\bibinfo  {journal} {ACM Transactions on Quantum Computing}\ }\textbf {\bibinfo {volume} {3}} (\bibinfo {year} {2022})}\BibitemShut {NoStop}%
\bibitem [{\citenamefont {Antipov}\ \emph {et~al.}(2023)\citenamefont {Antipov}, \citenamefont {Kiktenko},\ and\ \citenamefont {Fedorov}}]{Antipov_Kiktenko_Fedorov_2023}%
  \BibitemOpen
  \bibfield  {author} {\bibinfo {author} {\bibfnamefont {A.~V.}\ \bibnamefont {Antipov}}, \bibinfo {author} {\bibfnamefont {E.~O.}\ \bibnamefont {Kiktenko}},\ and\ \bibinfo {author} {\bibfnamefont {A.~K.}\ \bibnamefont {Fedorov}},\ }\bibfield  {title} {\bibinfo {title} {Realizing a class of stabilizer quantum error correction codes using a single ancilla and circular connectivity},\ }\href {https://doi.org/10.1103/PhysRevA.107.032403} {\bibfield  {journal} {\bibinfo  {journal} {Phys. Rev. A}\ }\textbf {\bibinfo {volume} {107}},\ \bibinfo {pages} {032403} (\bibinfo {year} {2023})}\BibitemShut {NoStop}%
\bibitem [{\citenamefont {Criger}\ and\ \citenamefont {Ashraf}(2018)}]{Criger_Ashraf_2018}%
  \BibitemOpen
  \bibfield  {author} {\bibinfo {author} {\bibfnamefont {B.}~\bibnamefont {Criger}}\ and\ \bibinfo {author} {\bibfnamefont {I.}~\bibnamefont {Ashraf}},\ }\bibfield  {title} {\bibinfo {title} {Multi-path summation for decoding 2d topological codes},\ }\href {https://doi.org/10.22331/q-2018-10-19-102} {\bibfield  {journal} {\bibinfo  {journal} {Quantum}\ }\textbf {\bibinfo {volume} {2}},\ \bibinfo {pages} {102} (\bibinfo {year} {2018})}\BibitemShut {NoStop}%
\bibitem [{\citenamefont {Roffe}\ \emph {et~al.}(2020)\citenamefont {Roffe}, \citenamefont {White}, \citenamefont {Burton},\ and\ \citenamefont {Campbell}}]{roffe_decoding_2020}%
  \BibitemOpen
  \bibfield  {author} {\bibinfo {author} {\bibfnamefont {J.}~\bibnamefont {Roffe}}, \bibinfo {author} {\bibfnamefont {D.~R.}\ \bibnamefont {White}}, \bibinfo {author} {\bibfnamefont {S.}~\bibnamefont {Burton}},\ and\ \bibinfo {author} {\bibfnamefont {E.}~\bibnamefont {Campbell}},\ }\bibfield  {title} {\bibinfo {title} {Decoding across the quantum low-density parity-check code landscape},\ }\bibfield  {journal} {\bibinfo  {journal} {Physical Review Research}\ }\textbf {\bibinfo {volume} {2}},\ \href {https://doi.org/10.1103/physrevresearch.2.043423} {10.1103/physrevresearch.2.043423} (\bibinfo {year} {2020})\BibitemShut {NoStop}%
\bibitem [{\citenamefont {Higgott}\ \emph {et~al.}(2023)\citenamefont {Higgott}, \citenamefont {Bohdanowicz}, \citenamefont {Kubica}, \citenamefont {Flammia},\ and\ \citenamefont {Campbell}}]{Higgott_Bohdanowicz_Kubica_Flammia_Campbell_2023}%
  \BibitemOpen
  \bibfield  {author} {\bibinfo {author} {\bibfnamefont {O.}~\bibnamefont {Higgott}}, \bibinfo {author} {\bibfnamefont {T.~C.}\ \bibnamefont {Bohdanowicz}}, \bibinfo {author} {\bibfnamefont {A.}~\bibnamefont {Kubica}}, \bibinfo {author} {\bibfnamefont {S.~T.}\ \bibnamefont {Flammia}},\ and\ \bibinfo {author} {\bibfnamefont {E.~T.}\ \bibnamefont {Campbell}},\ }\bibfield  {title} {\bibinfo {title} {Improved decoding of circuit noise and fragile boundaries of tailored surface codes},\ }\href {https://doi.org/10.1103/PhysRevX.13.031007} {\bibfield  {journal} {\bibinfo  {journal} {Physical Review X}\ }\textbf {\bibinfo {volume} {13}},\ \bibinfo {pages} {031007} (\bibinfo {year} {2023})}\BibitemShut {NoStop}%
\bibitem [{\citenamefont {Developers}(2024)}]{Developers_2024}%
  \BibitemOpen
  \bibfield  {author} {\bibinfo {author} {\bibfnamefont {C.}~\bibnamefont {Developers}},\ }\href {https://doi.org/10.5281/zenodo.11398048} {\bibinfo {title} {Cirq}} (\bibinfo {year} {2024})\BibitemShut {NoStop}%
\bibitem [{\citenamefont {Keppens}(2025)}]{Keppens2025}%
  \BibitemOpen
  \bibfield  {author} {\bibinfo {author} {\bibfnamefont {J.}~\bibnamefont {Keppens}},\ }\href {https://github.com/Keppens/5QuditCodeTutorial} {\bibinfo {title} {5quditcodetutorial}} (\bibinfo {year} {2025}),\ \bibinfo {note} {accessed: 2025-07-31}\BibitemShut {NoStop}%
\bibitem [{\citenamefont {Wilde}(2013)}]{Wilde_2013}%
  \BibitemOpen
  \bibfield  {author} {\bibinfo {author} {\bibfnamefont {M.~M.}\ \bibnamefont {Wilde}},\ }\href {https://doi.org/10.1017/CBO9781139525343} {\emph {\bibinfo {title} {Quantum Information Theory}}},\ \bibinfo {edition} {1st}\ ed.\ (\bibinfo  {publisher} {Cambridge University Press},\ \bibinfo {year} {2013})\BibitemShut {NoStop}%
\bibitem [{\citenamefont {Chao}\ and\ \citenamefont {Reichardt}(2020)}]{Chao_Reichardt_2020}%
  \BibitemOpen
  \bibfield  {author} {\bibinfo {author} {\bibfnamefont {R.}~\bibnamefont {Chao}}\ and\ \bibinfo {author} {\bibfnamefont {B.~W.}\ \bibnamefont {Reichardt}},\ }\bibfield  {title} {\bibinfo {title} {Flag fault-tolerant error correction for any stabilizer code},\ }\href {https://doi.org/10.1103/PRXQuantum.1.010302} {\bibfield  {journal} {\bibinfo  {journal} {PRX Quantum}\ }\textbf {\bibinfo {volume} {1}},\ \bibinfo {pages} {010302} (\bibinfo {year} {2020})}\BibitemShut {NoStop}%
\bibitem [{\citenamefont {Calderbank}\ \emph {et~al.}(1997)\citenamefont {Calderbank}, \citenamefont {Rains}, \citenamefont {Shor},\ and\ \citenamefont {Sloane}}]{Calderbank_Rains_Shor_Sloane_1997}%
  \BibitemOpen
  \bibfield  {author} {\bibinfo {author} {\bibfnamefont {A.~R.}\ \bibnamefont {Calderbank}}, \bibinfo {author} {\bibfnamefont {E.~M.}\ \bibnamefont {Rains}}, \bibinfo {author} {\bibfnamefont {P.~W.}\ \bibnamefont {Shor}},\ and\ \bibinfo {author} {\bibfnamefont {N.~J.~A.}\ \bibnamefont {Sloane}},\ }\href {https://doi.org/10.48550/arXiv.quant-ph/9608006} {\bibinfo {title} {Quantum error correction via codes over gf(4)}} (\bibinfo {year} {1997})\BibitemShut {NoStop}%
\bibitem [{\citenamefont {Yoshida}\ \emph {et~al.}(2025)\citenamefont {Yoshida}, \citenamefont {Tamiya},\ and\ \citenamefont {Yamasaki}}]{Yoshida_Tamiya_Yamasaki_2025}%
  \BibitemOpen
  \bibfield  {author} {\bibinfo {author} {\bibfnamefont {S.}~\bibnamefont {Yoshida}}, \bibinfo {author} {\bibfnamefont {S.}~\bibnamefont {Tamiya}},\ and\ \bibinfo {author} {\bibfnamefont {H.}~\bibnamefont {Yamasaki}},\ }\href {https://doi.org/10.48550/arXiv.2402.09606} {\bibinfo {title} {Concatenate codes, save qubits}} (\bibinfo {year} {2025}),\ \bibinfo {note} {arXiv:2402.09606 [quant-ph]}\BibitemShut {NoStop}%
\bibitem [{\citenamefont {Gidney}(2021)}]{gidney2021stim}%
  \BibitemOpen
  \bibfield  {author} {\bibinfo {author} {\bibfnamefont {C.}~\bibnamefont {Gidney}},\ }\bibfield  {title} {\bibinfo {title} {Stim: a fast stabilizer circuit simulator},\ }\href {https://doi.org/10.22331/q-2021-07-06-497} {\bibfield  {journal} {\bibinfo  {journal} {{Quantum}}\ }\textbf {\bibinfo {volume} {5}},\ \bibinfo {pages} {497} (\bibinfo {year} {2021})}\BibitemShut {NoStop}%
\end{thebibliography}%
\end{document}